\documentclass[11pt]{article}

\usepackage{appendix}
\usepackage{amssymb,amsmath,amsfonts,dsfont}
\usepackage{graphicx}
\usepackage[figuresright]{rotating}
\usepackage[applemac]{inputenc}
\usepackage{graphicx}
\usepackage{dcolumn}
\usepackage{bm}
\usepackage{amscd,amsthm}
\usepackage{ifthen}
\usepackage{booktabs}
\usepackage{bm}
\usepackage{placeins}
\usepackage{url}

\usepackage[skip=0pt]{caption}

\usepackage{hyperref}
\hypersetup{
	bookmarks=true,         
	unicode=false,          
	pdftoolbar=true,        
	pdfmenubar=true,        
	pdffitwindow=false,     
	pdfstartview={FitH},    
	pdfauthor={Johannes~F.~Kunz},     
	pdfsubject={Implicit Neural Networks with Fourier-Feature Inputs for Free-breathing Cardiac MRI Reconstruction},   
	pdfcreator={Johannes~F.~Kunz},   
	pdfproducer={Producer}, 
	pdfnewwindow=true,      
	colorlinks=true,       
	linkcolor=red,          
	citecolor=green,        
	filecolor=magenta,      
	urlcolor=cyan           
}
\usepackage{breakurl}

\usepackage[
maxbibnames=9,giveninits=true,style=alphabetic
]{biblatex}
\bibliography{main.bib}

\usepackage{tikz}
\usepackage{mathtools}
\usepackage{pgfplots}
\usepgfplotslibrary{groupplots} 
\usetikzlibrary{pgfplots.groupplots}
\usetikzlibrary{matrix,arrows,decorations.pathmorphing}
\usepackage{pgfplots,pgfplotstable}
\usepgfplotslibrary{fillbetween}
\usetikzlibrary{shadows,arrows}
\usetikzlibrary{positioning}

\usepgfplotslibrary{colorbrewer}
\pgfplotsset{compat=newest}
\usetikzlibrary{external}

\usepackage{graphics}
\usepackage{comment}
\usepackage{readarray} 
\usepackage{caption} 
\usepackage{changepage}

\usepackage{hyperref}       
\usepackage{url}            
\usepackage{amsfonts}       
\usepackage{nicefrac}       
\usepackage{microtype}      

\newdimen\nodeDist
\usepackage{url}
\usepackage{breakurl}
\usepackage{outlines} 
\usepackage{readarray} 
\usepackage{changepage} 

\usepackage[binary-units=true]{siunitx}
\usepackage{rotating}
\usepackage{multirow}
\usepackage{caption}
\usepackage{subcaption}
\usepackage{float}
\usepackage{calc}

\usepackage{amsmath}
\usepackage{amssymb}
\usepackage{bm}

\newcommand\realnumbers[0]{\mathbb{R}}

\newcommand\norm[1]{\left\Vert #1 \right\Vert}

\newcommand\meanop[1]{\mathbb{E}} 
\newcommand\normpdf[2]{\mathcal{N}\left(#1,\, #2\right)} 

\newcommand\meas[0]{\boldsymbol{y}}

\newcommand\state[0]{\sigma}
\newcommand\img[1][]{\boldsymbol{x}(\state_{#1})}
\newcommand\imgrec[1] [] {\hat{\boldsymbol{x}}(\state_{#1})}

\newcommand\measop[0]{\boldsymbol{A}}
\newcommand\mask[0]{\boldsymbol{M}}
\newcommand\smaps[0]{\boldsymbol{S}}
\newcommand\fft[0]{\boldsymbol{F}}

\newcommand\noise[0]{\boldsymbol{n}}

\newcommand\frameindex[0]{\tau}
\newcommand\frames[0]{T}

\newcommand\ky[0]{k_{\text{y}}}
\newcommand\kx[0]{k_{\text{x}}}

\newcommand\measrec[0]{\hat{\meas}} 

\newcommand\loss[0]{\mathcal{L}}

\newcommand\ser[0]{\text{SER}}

\newcommand\numlines[0]{N_{\text{lines}}}

\newcommand\coordinate[0]{\boldsymbol{c}}
\newcommand\fmlpparams[0]{\bm \theta}
\newcommand\fmlp[0]{f_{\fmlpparams}}
\newcommand\fmlpgrid[1][]{\boldsymbol{f}_{\fmlpparams #1}}
\newcommand\fmap[0]{\boldsymbol{\gamma}}
\newcommand\fmapwavevector[0]{\boldsymbol{b}}
\newcommand\fmapwavematrix[0]{\boldsymbol{B}}
\newcommand\csx[0]{s_{\text{x}}}
\newcommand\csy[0]{s_{\text{y}}}
\newcommand\cst[0]{s_{\text{t}}}
\newcommand\sout[0]{s_{\text{out}}}
\newcommand\hiddenlayers[0]{N_{\text{hidden}}}
\newcommand\imgrecold[0]{\hat{\boldsymbol{x}}}

\newcommand\sg[0]{\operatorname{sg}}
\newcommand\hdroffset[0]{\varepsilon}
\newcommand\lambdadenoiser[0]{\lambda_{\text{denoiser}}}
\newcommand\kfmlptrajectory[1][]{\fmlpgrid[#1]^{\text{traj}}}

\newcommand\nik[0]{f_{\fmlpparams}}
\newcommand\nikevaluation[1][]{\fmlpgrid[#1]}

\newcommand\imgwidthsamplingpattern[0]{3.9cm}
\newcommand\imgwidthreconstructions[0]{4.2cm}
\newcommand\figwidthfull[0]{11cm}
\newcommand\figheightfull[0]{8cm}
\newcommand\figwidthhalf[0]{\textwidth}

\begin{document}

\begin{center}
	
	{\bf{\LARGE{
				Implicit Neural Networks with Fourier-Feature Inputs for Free-breathing Cardiac MRI Reconstruction
	}}}
	
	\vspace*{.2in}
	
	{\large{
			\begin{tabular}{cccc}
				Johannes~F.~Kunz$^\ast$, Stefan~Ruschke$^\dagger$ , Reinhard~Heckel$^\ast$
			\end{tabular}
	}}
	
	\vspace*{.05in}
	
	\begin{tabular}{c}
	$^\ast$Dept. of Electrical and Computer Engineering, Technical University of Munich \\
	$^\dagger$ Dept. of Diagnostic and Interventional Radiology, Klinikum rechts der Isar, \\School of Medicine, Technical University of Munich 
	\end{tabular}

	\vspace*{.1in}

	\today
	
	\vspace*{.1in}
	
\end{center}

\begin{abstract}
Cardiac magnetic resonance imaging (MRI) requires reconstructing a real-time video of a beating heart from continuous highly under-sampled measurements. This task is challenging since the object to be reconstructed (the heart) is continuously changing during signal acquisition. In this paper, we propose a reconstruction approach based on representing the beating heart with an implicit neural network and fitting the network so that the representation of the heart is consistent with the measurements. 
The network in the form of a multi-layer perceptron with Fourier-feature inputs acts as an effective signal prior and enables adjusting the regularization strength in both the spatial and temporal dimensions of the signal.
We study the proposed approach for 2D free-breathing cardiac real-time MRI in different operating regimes, i.e., for different image resolutions, slice thicknesses, and acquisition lengths. Our method achieves reconstruction quality on par with or slightly better than state-of-the-art untrained convolutional neural networks and superior image quality compared to a recent method that fits an implicit representation directly to Fourier-domain measurements. 
However, this comes at a relatively high computational cost. 
Our approach does not require any additional patient data or biosensors including electrocardiography, making it potentially applicable in a wide range of clinical scenarios. 

\end{abstract}
\section{Introduction}
\label{sec:introduction}
Real-time magnetic resonance imaging (MRI) is a dynamic imaging technique for assessing the anatomic structure and function of moving organs in the human body.
The reconstruction of cardiac function is challenging due to the relatively fast organ movements compared to the achievable encoding speed.
In this paper, we consider the reconstruction of free-breathing cardiac real-time MRI data, where the goal is to reconstruct a video of the beating heart from highly under-sampled data continuously acquired over multiple cardiac and respiratory cycles. The video depicts the cardiac function in real-time meaning that arrhythmia and unprecedented motion can be imaged without assuming periodicity of the motion. 
This task is especially difficult when only very few samples are collected at any given time instance during the cardiac and respiratory cycles.

Various reconstruction methods for cardiac MRI are based on gated or triggered data \cite{lanzerCardiacImagingUsing1984,larsonSelfGatedCardiacCine2004}. Such methods bin the measurements according to different times in the cardiac and respiratory cycle. The bins are reconstructed independently using methods for static MRI, or they are reconstructed jointly using signal priors for regularizing among bins \cite{fengXDGRASPGoldenangleRadial2016,christodoulouMagneticResonanceMultitasking2018,jiangMotionRobustHigh2018}. However, binning can lead to artifacts since the heart may not be exactly in the same position at a given point in the cardiac and respiratory cycle. 

Another class of methods reconstructs images from continuously acquired un-gated data, without relying on gating or triggering, by regularizing in the spatial and temporal dimensions with total-variation norm penalties~\cite{fengGoldenangleRadialSparse2014}, through imposing low-rank priors~\cite{ongExtremeMRILargeScale2020,nakarmiKernelbasedLowrankKLR2017,poddarManifoldRecoveryUsing2019,poddarManifoldRecoveryUsing2019}, or both \cite{lingalaAcceleratedDynamicMRI2011}. Thereby, the heart can be imaged in real-time without binning errors.

With the advent of deep learning, signal priors can be learned from data. Supervised methods have been proposed for dynamic MRI that are trained on pairs of measurement data and corresponding ground-truth images~\cite{schlemperDeepCascadeConvolutional2017, qinConvolutionalRecurrentNeural2019, kustnerCINENetDeepLearningbased2020, huangDeepLowrankSparse2021, qin2021, huangDeepLowrankSparse2021}. However, ground-truth fully sampled data for real-time MRI is generally not available without assuming periodicity of the motion.

A recent elegant approach to avoid the use of training data but benefit from the prior inherent in neural networks are untrained networks that regularize by fitting a neural network to the measurement data. Untrained neural networks are typically based on convolutional neural networks (CNNs) \cite{yooTimeDependentDeepImage2021, zouDynamicImagingUsing2021, ahmedDynamicImagingUsing2022, biswasDynamicMRIUsing2019}, since CNNs have a provable bias towards fitting smooth signals~\cite{heckelDenoisingRegularizationExploiting2020, heckelCompressiveSensingUntrained2020}. 

In this paper, we propose an approach for real-time cardiac MRI that is based on representing the beating heart with a multi-layer perceptron (MLP) with Fourier-feature inputs and fitting the network so that the representation is consistent with the measurements. Our approach is untrained, i.e., it does not rely on any training data. 
Using a Fourier-feature MLP is critical for our approach since it enables us to impose a strong spatial and temporal prior on the heart to be reconstructed. 
 The method enables control over the strength of spatial and temporal regularization by setting up the Fourier-feature inputs.



We compare our approach to the time-dependent deep image prior (t-DIP) \cite{yooTimeDependentDeepImage2021}, a state-of-the-art untrained method based on CNNs. 
We study the reconstruction quality and computational cost on various datasets in different operating regimes, including different image resolutions, slice thicknesses, and acquisition lengths, and find our method performs on par in terms of image quality, sometimes even marginally better, but at a higher computational cost. We thereby are the first to show that an untrained reconstruction method based on implicit networks can perform on-par or outperform state-of-the-art CNN-based untrained reconstruction methods, despite the very different network structure.

Compared to two recently proposed approaches~\cite{huangNeuralImplicitKSpace2022, fengSpatiotemporalImplicitNeural2023} also based on implicit neural representations, we achieve better image quality. 

Specifically, Huang et al.~\cite{huangNeuralImplicitKSpace2022} enable cardiac MRI by fitting a Fourier-feature MLP directly to the measurements in the Fourier domain. Fitting the MLP in the Fourier domain does not require temporal binning of measurements and circumvents non-uniform Fourier transforms in case of non-Cartesian sampling patterns, as opposed  to fitting it in the image domain. However, this comes at a significant loss in image quality, as we show later, since Fourier-feature MLPs are not a good model for representing an image in the Fourier domain. 
Specifically, an image is relatively smooth, whereas the spectrum of an image is not smooth, and a Fourier-feature MLP has a bias towards fitting smooth signals.
Since Huang et al.'s method is computationally intensive, we also consider a variant of the method with substantially shorter training time.

Feng et al.~\cite{fengSpatiotemporalImplicitNeural2023}'s method is based on fitting an MLP with hash encoding to the data. Using a hash encoding also has computational benefits over our approach, but a hash encoding is not an effective image prior, thus the method requires additional explicit regularization.


\section{Related Work}
\label{sec:related-work}

In this section, we briefly discuss classical approaches to real-time cardiac MRI, implicit neural networks, and CNN-based neural networks for medical image reconstruction. 

\paragraph*{\bf Approaches to cardiac MRI with sparse and low-rank regularization techniques}
A variety of classical approaches to cardiac MRI are based on regularization in the spatial and temporal dimensions with hand-crafted signal priors. 
For example, Feng et al.~\cite{fengGoldenangleRadialSparse2014} impose a total-variation norm penalty over time and Feng et al.~\cite{fengXDGRASPGoldenangleRadial2016} impose a total-variation penalty over cardiac phases. 

Another line of work incorporates temporal relations by assuming low-rank structures within the temporal series of images through low-rank tensor decompositions \cite{christodoulouMagneticResonanceMultitasking2018}, 
through decompositions into image patches of low-rank \cite{jiangMotionRobustHigh2018}, and through multi-scale low-rank decompositions \cite{ongExtremeMRILargeScale2020}.

Besides these models, there are methods that assume low-rankness in a kernel space \cite{nakarmiKernelbasedLowrankKLR2017, poddarManifoldRecoveryUsing2019} or assume that the images lie on a smooth low-dimensional manifold \cite{poddarDynamicMRIUsing2016, ahmedFreebreathingUngatedDynamic2020}.

\paragraph*{\bf Implicit neural networks}
Our work relies on representing time-varying signals with an implicit neural network, i.e., a function parameterized by a neural network that maps a spatial coordinate vector and a time variable to a pixel value. 
Implicit neural networks are used for representing images and volumes in a variety of applications. 
The architecture of the network is important for images and volumes transforming the coordinates with a Fourier-feature map before passing them through a multi-layer perceptron works well, and most current architectures including NeRF~\cite{mildenhallNeRFRepresentingScenes2020a}, SIREN~\cite{sitzmannImplicitNeuralRepresentations2020}, and MLPs with Fourier-feature inputs~\cite{tancikFourierFeaturesLet2020} use this architecture. 

\paragraph*{\bf Implicit networks for medical imaging}
Implicit neural networks are excellent image models and can therefore be used as an image prior for image reconstruction tasks, as demonstrated by Tancik et al.~\cite{tancikFourierFeaturesLet2020} on a toy dataset. 
Subsequently, 
Shen et al.~\cite{shenNeRPImplicitNeural2021} have improved the reconstruction quality for static CT and MRI imaging by pre-training the network on a previously reconstructed image of the same subject. 
Shen et al.~\cite{shenNeRPImplicitNeural2021} also reconstructed a temporal series of 3D images and used the first image for pre-training the network. Contrary to our approach, the method does not model temporal variations. 

Implicit networks have also been used for super-resolution tasks in medical imaging, for obtaining high-resolution 3D MRI~\cite{wuIREMHighResolutionMagnetic2021} from low-resolution 2D images, and for building a scale-agnostic model for MRI~\cite{vanveenScaleAgnosticSuperResolutionMRI2022}. 

\paragraph*{\bf Implicit networks for dynamic MRI reconstruction}
As mentioned before, most related to our work are two recent methods using implicit networks for dynamic MRI reconstruction. 
Huang et al.~\cite{huangNeuralImplicitKSpace2022} introduced neural implicit k-space (NIK), which uses a SIREN network \cite{sitzmannImplicitNeuralRepresentations2020} with additional explicit regularization to represent the signal in k-space. It outputs a k-space value given the k-space coordinate, the time, and the receiver coil index. Once the network has been trained on the measured frequencies, it can be evaluated at the missing k-space frequencies to yield a reconstructed k-space.
We compare our approach to the NIK method.

Feng et al. \cite{fengSpatiotemporalImplicitNeural2023} learn a neural representation of the dynamic object in the image domain in the form of an MLP with a hash-encoding of the input coordinates. The MLP is small compared to a Fourier-feature MLP and can be evaluated efficiently. However, the method relies on additional explicit regularization of the images' temporal variation and an additional loss term to enforce the low-rankness of the solution, as opposed to only an implicit bias inherent to the network architecture as we do.

\paragraph*{\bf Untrained methods based on CNNs}
Our approach relies on regularization with an MLP with Fourier-feature inputs. We find that the network has a regularizing effect that is similar to state-of-the-art untrained CNNs.

Untrained neural networks introduced by Ulyanov et al. in the DIP paper~\cite{ulyanovDeepImagePrior2020} allow image reconstruction by fitting a neural network to the measurement data. Simple CNNs provide excellent signal reconstruction performance without any training~\cite{heckelDeepDecoderConcise2019}, and provably reconstruct smooth signals~\cite{heckelDenoisingRegularizationExploiting2020, heckelCompressiveSensingUntrained2020}. 
Untrained networks can outperform classical sparsity-based reconstruction for accelerated MRI~\cite{darestaniAcceleratedMRIUntrained2021}. 

The time-dependent deep image prior (t-DIP) \cite{yooTimeDependentDeepImage2021} extends the deep image prior to dynamic MRI.
Like our approach, t-DIP fits a single model to the entire series of images, and time is used as an input coordinate.
In contrast to using a CNN to regularize in the spatial dimension, we model both the temporal and the spatial dimensions with an implicit network. 

The method Gen-SToRM \cite{zouDynamicImagingUsing2021} trains a CNN to generate a temporal series of images from a temporal series of low-dimensional network inputs. 
A conceptually similar method has been applied for video reconstruction \cite{hyderGenerativeModelsLowDimensional2020}.

\section{Problem Formulation}
\label{sec:problem-formulation}

Our goal is to reconstruct a video of a moving object from sequentially acquired, under-sampled linear measurements. 
We consider an object $\img \in \mathbb C^{w\times h}$ in the form of an image of width $w$ and height $h$ that is parameterized by a motion state $\state$. In our setup, the object $\img$ is the image of a heart at a certain state in the cardiac and respiratory cycle.

We consider the standard MRI measurement model, where we collect noisy linear measurements 
\begin{align}
\label{eq:measurementmulticoil}
    \meas_{\frameindex,c} = \mask_{\frameindex}\fft\smaps_{c}\img[\frameindex] + \noise_{\frameindex,c}
\end{align}
at the receiver coils $c = 1, \dots, C$. 
Here, $\fft$ is the 2-D Fourier matrix, $\smaps_{c}$ is a diagonal matrix containing the sensitivities of the receiver coil $c$, and $\mask_{\frameindex}$ is a binary sampling mask encoding the frequencies collected during time $\frameindex$. 
In this work, we consider Cartesian sampling patterns.
The measurements are distorted by additive noise $\noise_{\frameindex,c}$.

Note that the object $\img$ is changing continuously. The measurement model~\eqref{eq:measurementmulticoil} assumes that the object is essentially constant during a very short time frame, which we index by $\frameindex$. During this short time frame, only very few measurements, i.e., the frequencies selected by the mask $\mask_{\frameindex}$, are collected.

The measurements from different coils can be stacked into a single measurement vector
\begin{equation*}
    \meas_{\frameindex} = \measop_{\frameindex} \img[\frameindex] + \noise_{\frameindex},
\end{equation*}
where $\meas_{\frameindex} = [\meas_{\frameindex, 1}^T, \dots, \meas_{\frameindex, C}^T]^T$ and $\noise_{\frameindex} = [\noise_{\frameindex, 1}^T, \dots, \noise_{\frameindex, C}^T]^T$. The matrix $\measop_{\frameindex}$ is the resulting forward map for frame $\frameindex$.

Our goal is to reconstruct the image series (or video) $\img[\frameindex], \, \frameindex=1, \dots, \frames$ from the noisy measurements $\meas_{\frameindex}$. 
During each time frame indexed by $\frameindex$, only very few measurements are collected. Therefore, reconstructing each frame individually results in poor image quality, even if we take prior information about the image into account. Successful reconstruction relies on using prior information about the images in spatial directions as well as in the temporal direction. 

Note that in our problem formulation, the index $\frameindex$ refers to a short time frame that is used for binning the measurement data such that motion can be neglected. This time frame may not necessarily correspond to the time frame that is used in the video that is ultimately rendered. 
An advantage of our method based on implicit representation is that we can use different frame rates for reconstruction and for the output shown to the end-user.


\section{Methods}
\label{sec:method}

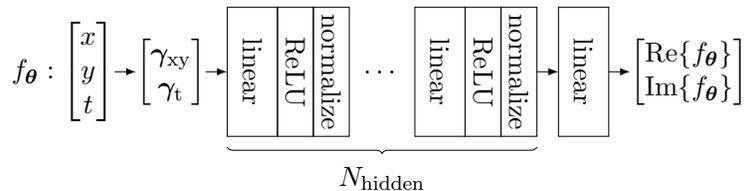
\begin{figure}[h]  
    \centering
    \resizebox{0.6\columnwidth}{!}{

 
 

\begin{tikzpicture}[every node/.style={inner sep=0,outer sep=0}]

\node [anchor=west] (input) at (0,0) {$\fmlp: \begin{bmatrix} x\\ y\\ t\end{bmatrix}$};
\node [
    right=0.3cm of input.east,
    anchor=west] (embedding) {$\begin{bmatrix}\fmap_{\text{xy}} \\ \fmap_{\text{t}}\end{bmatrix}$};

\node [
    draw,
    rectangle,
    anchor=west,
    minimum height=1.8cm,
    minimum width=0.7cm,
    right=0.3cm of embedding.east,
    ] (linear1) {\begin{turn}{-90}linear\end{turn}};
    
\node [
    draw,
    rectangle,
    anchor=south,
    minimum height=1.8cm,
    minimum width=0.5cm,
    right=0.7cm of linear1.west,
    ] (relu1) {\begin{turn}{-90}ReLU\end{turn}};
    
\node [
    draw,
    rectangle,
    anchor=east,
    minimum height=1.8cm,
    minimum width=0.5cm,
    right=0.5cm of relu1.west,
    ] (norm1) {\begin{turn}{-90}normalize\end{turn}};

\node [
    right=0.2cm of norm1.east,
    anchor=west] (dots) {$\dots$};

\node [
    draw,
    rectangle,
    anchor=west,
    minimum height=1.8cm,
    minimum width=0.7cm,
    right=0.2cm of dots.east,
    ] (linear2) {\begin{turn}{-90}linear\end{turn}};
    
\node [
    draw,
    rectangle,
    anchor=south,
    minimum height=1.8cm,
    minimum width=0.5cm,
    right=0.7cm of linear2.west,
    ] (relu2) {\begin{turn}{-90}ReLU\end{turn}};
    
\node [
    draw,
    rectangle,
    anchor=east,
    minimum height=1.8cm,
    minimum width=0.5cm,
    right=0.5cm of relu2.west,
    ] (norm2) {\begin{turn}{-90}normalize\end{turn}};

\node [
    draw,
    rectangle,
    anchor=west,
    minimum height=1.8cm,
    minimum width=0.7cm,
    right=0.3cm of norm2.east,
    ] (linear3) {\begin{turn}{-90}linear\end{turn}};

\node [
    right=0.3cm of linear3.east,
    anchor=west] (output) {$\begin{bmatrix}\operatorname{Re}\{\fmlp \} \\ \operatorname{Im}\{\fmlp \}\end{bmatrix}$};
    
\draw [draw, -latex] (input.east) -- (embedding.west);
\draw [draw, -latex] (embedding.east) -- (linear1.west);
\draw [draw, -latex] (norm2.east) -- (linear3.west);
\draw [draw, -latex] (linear3.east) -- (output.west);


\path (linear1.south west)
        edge [decorate, decoration={brace, mirror, raise=.15cm}]
        (linear1.south west  -| norm2.south east);
\node [
    below=1.3cm of dots.south,
    anchor=north] (output) {$N_{\text{hidden}}$};
    
\end{tikzpicture}
}
    \vspace{1em}
    \caption{The FMLPs' network consists of separate spatial and temporal Fourier-feature embeddings that are concatenated, followed by an MLP that outputs the complex image intensity at the specified coordinate.}
    \label{fig:fmlp-architecture}
\end{figure}

Our FMLP method is based on parameterizing the object $\img$ with a Fourier-feature MLP (FMLP) and fitting the MLP to the measurement data. Once fitted, we query the neural network to generate a video as an output. The Fourier-feature MLP we use is a good prior for a (slowly) time-varying smooth object, and we use no other regularization for reconstruction.

The object $\img$ is parameterized by a FMLP 
$\fmlp: \realnumbers^3 \rightarrow \realnumbers^{2}$ 
mapping a spatio-temporal coordinate vector $\coordinate = [x, y, t]^T$ to a complex image intensity value, represented by a real and an imaginary part. We focus on 2D images, but the method can easily be extended to 3D by adding another spatial coordinate so that $\coordinate = [x,y,z,t]$. 
The coordinates $x$ and $y$ denote physical locations within the field of view measured in meters, and the time-coordinate $t$ is time measured in seconds. The origins of the spatial and temporal coordinate axes can be set arbitrarily as the particular neural network that we use is shift-invariant. 

We set $t = \SI{0}{\second}$ as the beginning of the measurement process. For computing a pixel-based image $\fmlpgrid(t_{\frameindex})$ at time $t_{\frameindex}$, the network $\fmlp$ is evaluated on a regular grid of spatial coordinates within the field of view with the temporal coordinate fixed to $t_{\frameindex}$.

We fit a network by minimizing the reconstruction loss defined as
\begin{equation*}
    \loss(\fmlpparams) = \frac{1}{\frames} \sum\limits_{\frameindex=1}^{\frames} \norm{\measop_{\frameindex} \fmlpgrid(t_{\frameindex}) - \meas_{\frameindex}}_2^2.
\end{equation*}
The loss ensures data consistency of the images generated by the FMLP and the measurements. 
We minimize the loss with the Adam optimizer, and we take a batch size of one by choosing a frame uniformly at random at each step and computing the gradient with respect to the random frame. 
The time $t_{\frameindex}$ is chosen to be at the center of the acquisition time window of frame $\frameindex$ and the spatial coordinate grid used for computing $\fmlpgrid(t_{\frameindex})$ is matched to the Cartesian sampling grid in k-space.
After fitting, reconstructed images can be sampled from the implicit representation as $\imgrec[\frameindex] = \fmlpgrid(t_{\frameindex})$.

The network architecture is critical for performance. Our architecture is based on using Fourier-features as inputs to a ReLU-MLP and is depicted in Figure~\ref{fig:fmlp-architecture}. The Fourier-feature inputs are used in a variety of implicit neural network architectures~\cite{tancikFourierFeaturesLet2020, sitzmannImplicitNeuralRepresentations2020, mildenhallNeRFDarkHigh2022}. 
The Fourier-feature layer embeds the spatial coordinates and the temporal coordinate separately as $\fmap(\coordinate) = [\fmap_{\text{spatial}}([x,y]),\, \fmap_{\text{temporal}}(t)]^T$ with
\begin{align*}
    \fmap_{\text{spatial}}([x,y]) 
    &= \left[\sin\left(\fmapwavematrix\begin{bmatrix}\csx x\\ \csy y\end{bmatrix}\right),\, \cos\left(\fmapwavematrix\begin{bmatrix} \csx x \\ \csy y\end{bmatrix}\right) \right]^T, \\
    \fmap_{\text{temporal}}(t) 
    &= \left[\sin(\fmapwavevector \cst t),\, \cos(\fmapwavevector \cst t) \right]^T.
\end{align*}
Here, the $\sin$ and $\cos$ functions are applied element-wise. The coordinates are scaled by hyper-parameters $\csx$, $\csy$, and $\cst$ with units $\frac{1}{\SI{}{\metre}}$ and $\frac{1}{\SI{}{\second}}$, respectively, which control the variance of the angular frequencies in the spatial and temporal dimensions. The elements in the matrix $\fmapwavematrix$ and the vector $\fmapwavevector$ are drawn independently from a zero-mean Gaussian distribution with unit variance $\normpdf{0}{1}$ and are fixed and not optimized over. We choose $\fmapwavematrix \in \realnumbers^{256 \times 2}$ and $\fmapwavevector \in \realnumbers^{64}$ for a total feature size of $512 + 128$.

Embedding the spatial and temporal coordinates separately leads to sharper reconstructions compared to the intuitive setup of having one Fourier-feature embedding for all three coordinates together, as demonstrated in an ablation study in the code supplement. 

The MLP takes the Fourier-feature vector $\fmap(\coordinate)$ as input. After each linear hidden layer, the ReLU activation is applied and the output feature vector is normalized to zero mean and unit variance. The output layer is linear and maps to two real numbers corresponding to the real and imaginary parts of the complex image intensity value. The weights of the linear layers are initialized by drawing from $\normpdf{0}{\sigma_{\text{linear}}^2}$, where the variance $\sigma_{\text{linear}}^2$ can be tuned. The final output is multiplied by a constant $s_{\text{out}}$ that is tuned as a hyper-parameter depending on the scaling of the k-space data and the sensitivity maps. This is equivalent to normalizing the data and tuning a fixed normalization factor, since we use only a few datasets for our experiments. In our implementation, we set $\sigma_{\text{linear}} = 0.01$ and the MLP consists of 7 hidden layers with 512 neurons each if not explicitly stated otherwise.

\subsection{Baseline methods}

In our experimental results, we consider three methods for comparison, which are described in more detail below. The first method is Huang et al.~\cite{huangNeuralImplicitKSpace2022}'s NIK method, which is based on fitting a Fourier-feature SIREN network in the Fourier domain with receiver coil index as a fourth input dimension.

As the second method, we consider a variant of the NIK method that fits a Fourier-feature MLP with ReLU activation function similar to the FMLP, but in k-space. Hence we call it KFMLP. In contrast to the NIK, the KFMLP has no additional dimension for the coil index allowing faster training and evaluation. 

The third method we compare to is the time-dependent DIP \cite{yooTimeDependentDeepImage2021}, a state-of-the-art untrained method based on CNNs.

\subsubsection{Neural implicit k-space (NIK)}

The NIK, as presented by Huang et al. \cite{huangNeuralImplicitKSpace2022} parameterizes the time-varying k-space $\fft \smaps \img$ with a neural network. 
The network $\nik: \realnumbers^{4} \rightarrow \realnumbers^{2}$ is a SIREN network (8 layers, 512 neurons per layer) that maps frequency-time-coil coordinates $[\kx, \ky, t, c]$ to complex k-space values. All coordinates are normalized to $[-1, 1]$. The network is fitted to the measured k-space data at coordinates along the sampling trajectory. Each coordinate-value pair serves as a sample in the training dataset $\{([\kx{_{,n}}, \ky{_{,n}}, t_n, c_n], \meas_n)\}_{n=1,\dots,N}$ that is randomized and trained on in batches. 

The NIK uses a loss function consisting of a high dynamic range reconstruction loss and an explicit k-space regularization term
\begin{equation*}
    \loss(\fmlpparams) = \frac{1}{N}\sum\limits_{n=1}^{N}\loss_{\text{HDR}}(\nikevaluation[,n], \meas_{n}) + \lambdadenoiser R(\nikevaluation[,n]).
\end{equation*}
The high dynamic range loss is defined as
\begin{equation*}
    \loss_{\text{HDR}}(\measrec, \meas) = \norm{\frac{\measrec - \meas}{|\sg(\measrec)| + \hdroffset}}_2^2.
\end{equation*}
Here, the operator $\sg(\cdot)$ stops the propagation of the gradient during back-propagation, $|\cdot|$ computes the absolute value accounting for the positive and negative k-space values, and $\hdroffset > 0$ is a hyper-parameter that adjusts the compression of the dynamic range. The division is computed element-wise.

The NIK's k-space regularization term $R(\nikevaluation[,n])$ is defined as
\begin{equation*}
    R(\nikevaluation[,n]) =  \loss_{\text{HDR}}(\nikevaluation[,n], \boldsymbol{K}_{n}\nikevaluation[,n]),
\end{equation*}
where $\boldsymbol{K}_{n}$ is a diagonal matrix with entries $e^{-d/2\sigma^2}$ that weigh the k-space values according to their respective coordinate distances $d = \sqrt{\kx^2 + \ky^2}$ to the k-space center.


After training, a reconstructed image of at time $t_{\frameindex}$ is obtained by evaluating the network at all frequencies on the Cartesian grid and combining the reconstructed k-spaces via coil combination. 

\subsubsection{Fourier-feature MLP as implicit k-space prior (KFMLP)}
\label{subsec:kfmlp}
The KFMLP is similar to the NIK \cite{huangNeuralImplicitKSpace2022}, but uses a network $\fmlp: \realnumbers^{3} \rightarrow \realnumbers^{2C}$ that maps frequency-time coordinates $[\kx, \ky, t]$ to complex k-space values that are represented by a real and an imaginary part for each of the $C$ receiver coils. The frequency-coordinates $\kx$ and $\ky$ are normalized to $[-\pi, \pi)$ and the time-coordinate $t$, as for the FMLP, measures time in seconds. 

The KFMLP computes the k-space values for all receiver coils at a given frequency-time coordinate with a single forward pass; thus, evaluating the KFMLP is faster than evaluating the NIK that requires $C$ forward passes for $C$ coils.

For computational purposes, we segment the trajectory into frames that are treated as samples of the dataset. Specifically, we evaluate the network $\fmlp$ at the k-space coordinates and sampling times that correspond to the measurements in the vectors $\meas_{\frameindex}$. The resulting reconstructed k-space along the trajectory of frame $\frameindex$ is denoted as $\kfmlptrajectory[,\frameindex]$. 

We minimize the same loss as proposed for the NIK \cite{huangNeuralImplicitKSpace2022}.

The hyper-parameters $\lambdadenoiser > 0$ and $\sigma > 0$ adjust the regularization strength. By ablation studies, we find that $\hdroffset = 10^4$, $\sigma = 10$ and $\lambdadenoiser = 0.1$ yield good results for our datasets.
The high dynamic range reconstruction loss and the k-space regularization term are not essential, but yield a small improvement compared to using the standard $l_2$ reconstruction loss without any explicit k-space regularization, as demonstrated by an ablation study in the code supplement. 

The KFMLP uses the same network architecture as the FMLP with Fourier-features that are given as input to an MLP with ReLU activation function and normalization. The k-space coordinates are embedded in a separate Fourier-feature vector from the time coordinate. As before, the coordinate scales $\csx$, $\csy$, and $\cst$ are tuned as hyper-parameters. 
The KFMLP architecture differs from the NIK architecture that uses a sinusoidal activation function, jointly embeds the time and k-space coordinates, treats the coils $c=1, \dots, C$ as an additional input dimension, and does not tune the coordinate scales hyper-parameters.

\subsubsection{Time-dependent DIP}
\label{subsec:sota-methods}
The time-dependent DIP \cite{yooTimeDependentDeepImage2021} is a state-of-the-art untrained reconstruction method that uses a CNN with time-varying network inputs as implicit prior. The time-varying network inputs are generated by embedding a temporal input coordinate on a helix trajectory and then mapping the embedding to the CNN's input features using a small fully connected network.

The helix is a trajectory in $\realnumbers^3$ with unit radius and slope $z_{\text{slack}}$ that we tune as a hyper-parameter. The angular frequency of the helix introduces prior information about the heart rate and periodicity of motion. To match the angular frequency of the helix with the heart rate, an estimate of the number of cardiac cycles is required, which we obtain from an electrocardiogram (ECG). 

Since the validation lines are extracted at random positions, the frames are not equally spaced in time. We account for this by modifying the sampled positions on the helix trajectory. Instead of sampling equally spaced points based on the frame index $k$ as proposed by the authors, we use the measurement time $t_{\frameindex}$ to ensure that the extraction of validation lines does not diminish the performance of the t-DIP. The sampled positions on the helix are mapped to a more expressive latent space by a small MLP called MapNet. The output of the MapNet is reshaped as input for the subsequent CNN. 

We modify the original CNN architecture to fit our image resolution.
The proposed architecture of the CNN up-samples the feature resolution in powers of two using nearest-neighbor interpolation. We adjust the input feature resolution and the number of up-sampling layers to obtain an output resolution that is close to our target resolution. The output image is then center-cropped to the target resolution. Two convolutional layers are applied between interpolation layers. We tune the channel depth of the convolutional layers as a hyper-parameter.


\section{Experiments and Results}
\label{sec:experiments}
We evaluate the reconstruction quality and computational cost of FMLP using the measurement data we collected. We compare our approach to the NIK, KFMLP, and the time-dependent DIP, and find that the FMLP performs as well as the t-DIP and better than the NIK and KFMLP in terms of image quality, but at a higher computational cost. Code to reproduce the results is available at \href{https://github.com/MLI-lab/cinemri}{https://github.com/MLI-lab/cinemri}.

\subsection{Datasets}
\label{subsec:datasets}
We conduct experiments on datasets that were acquired of a healthy 30-year-old male volunteer on a \SI{3}{\tesla} Elition X scanner (Philips Healthcare, The Netherlands). This study was approved by the local ethics committee and written informed consent was obtained. The datasets are available online \cite{datasets}.

Measurement data was acquired at two different resolutions 2.27 x 2.26$\SI[]{}{\milli\metre^2}$ ($264 \times 186$ acquisition matrix size) and 1.25 x 1.26$\SI[]{}{\milli\metre^2}$ ($480 \times 334$), with $\SI{10}{\milli \metre}$ and $\SI{5}{\milli \metre}$ slice thickness, respectively. We refer to those as \textbf{low-resolution high-SNR} and \textbf{high-resolution} datasets. 
A third dataset, referred to as \textbf{low-resolution low-SNR}, was acquired at low resolution with a reduced (isotropic) slice thickness of $\SI{2.25}{\milli \metre}$, decreasing the signal strength and the signal-to-noise ratio (SNR).  
For each configuration, a breath-hold triggered scan and a free-breathing scan were taken.
The acquisition length was set retrospectively for the free-breathing scan and we conduct experiments with $\SI{4.0}{\second}$, $\SI{8.0}{\second}$, and $\SI{15.9}{\second}$ of acquired data.

For the breath-hold scan, the data acquisition is triggered by an ECG such that measurements of similar cardiac phases are binned. 
The dynamics are then reconstructed with a sparsity-based method for static MRI which the MRI scanner performs by default.
The breath-hold reconstructions serve as a visual reference for the free-breathing reconstructions.

During the free-breathing scan, measurements are acquired continuously (ungated). We use a partial-Fourier Cartesian sampling pattern, where the k-space is fully sampled along the $\kx$-dimension (frequency-encoding) and randomly under-sampled along the $\ky$-dimension (phase-encoding). Thus, measurements are taken sequentially along $\ky$-lines in the k-space. The measured $\ky$-lines are binned into frames retrospectively.\footnote{Further details on the measurement parameters are available in the supplements of the released code.}

The sensitivity maps used for the reconstruction are estimated from separate calibration scans by the scanner using ESPIRiT. The free-breathing (FB) scans are taken with $C=25$ active receiver coils, while the breath-hold (BH) scans use a total of $C=26$ receiver coils. No further coil compression techniques are applied.

\subsection{Performance metrics and quality criteria}
\label{subsec:metrics}

For our measurement data (and original real-time MRI data in general), ground-truth images are not available. We therefore measure performance in terms of an estimate of the MSE as well as through visual comparisons.

We visually compare the reconstructed images to the ECG-triggered breath-hold reconstructions obtained from the scanner by binning-based reconstruction methods using sparsity regularization. We focus on anatomic details with the region of interest, the heart, such as papillary muscles in the left ventricle that are rapidly changing shape during systoles.

We estimate a normalized version of the mean squared error based on a hold-out set of randomly subsampled frequencies as follows. 
We randomly extract \SI{5}{\percent} of the measured $\ky$-lines for validation. The remaining $\ky$-lines are binned into frames $\frameindex = 1,\dots, \frames$ for training, such that the k-space of every frame contains $\numlines = 6$ many $\ky$-lines. Each validation line $\meas_{v}, v = 1, \dots, V$ is assigned to the frame $\frameindex_v$ that is closest in sample time. 
Let $\measrec_v$ be the reconstructed k-space lines predicted by a reconstruction algorithm (FMLP, KFMLP, and t-DIP in our setup). The lines are obtained by first applying the sensitivity maps and Fourier transform to the reconstructed images $\imgrec[\frameindex_v]$ and then extracting the $\ky$-line corresponding to $\meas_{v}$. 
We estimate the signal-to-error ratio (SER) as a metric for quantifying data consistency
\begin{equation*}
    \ser = 10 \log_{10} \frac{\sum\limits_{v = 1}^{V} \norm{\meas_{v}}_2^2}{\sum\limits_{v = 1}^{V} \norm{\measrec_{v} - \meas_{v}}_2^2}.
\end{equation*}

\subsection{Implementation details}
\label{subsec:implementation-details}
We implemented the FMLP, KFMLP, and t-DIP using PyTorch, and used the NIK implementation by Huang et al. \cite{huangNeuralImplicitKSpace2022}.
The models are optimized with Adam 
until no new SER high score has been reached for 200 epochs. We evaluate the reconstruction quality at the epoch of maximum SER. 
The experiments are conducted on a server equipped with a RTX 6000 GPU with \SI{24}{\giga \byte} VRAM and an Intel(R) Core(TM) i9-9940X CPU with \SI{128}{\giga \byte} available RAM.

\subsection{Quantitative reconstruction quality}
\label{subsec:quantitive-analyis}
\label{subsec:data-consistency}

In this section, we quantitatively compare the reconstruction quality of the FMLP, the NIK, the KFMLP, and the t-DIP in terms of the estimated normalized MSE (specifically, in terms of the SER defined in the previous section).

\paragraph*{\bf Different operating regimes}
We start by comparing the image reconstruction quality measured in terms of the SER in different operating regimes that feature different image resolutions and slice thicknesses. Specifically, we fit all three methods to the low-resolution high-SNR dataset, the low-resolution low-SNR dataset, and the high-resolution dataset. The hyper-parameters of the models are tuned on each dataset individually and the methods are trained on $\frames=225$ frames.

The results in Table~\ref{tab:ser-datasets} show that the FMLP and the t-DIP achieve a similar SER with marginal differences in all evaluated operating regimes. While the t-DIP achieves a slightly higher SER than the FMLP on the low-resolution high-SNR dataset, the FMLP performs slightly better on the low-resolution low-SNR and the high-resolution datasets. The NIK and the KFMLP, learning a representation of the k-space, achieve much lower SER scores than the FMLP or the t-DIP. The difference is substantial on all evaluated datasets.

\paragraph*{\bf Performance as a function of acquisition length}
Next, we study the reconstruction quality for different measurement acquisition lengths. 
We measure the image quality in terms of the SER within the first \SI{4}{\second} of reconstructed images, while training the methods on increasing acquisition lengths, i.e., on \SI{4}{\second}, \SI{8}{\second}, and \SI{16}{\second} of acquired data, respectively. The experiment is conducted on the low-resolution high-SNR dataset.

\begin{figure}[]
    \centering
\begin{tikzpicture}

\definecolor{crimson2143940}{RGB}{214,39,40}
\definecolor{darkgray176}{RGB}{176,176,176}
\definecolor{darkorange25512714}{RGB}{255,127,14}
\definecolor{forestgreen4416044}{RGB}{44,160,44}
\definecolor{lightgray204}{RGB}{204,204,204}
\definecolor{mediumpurple148103189}{RGB}{148,103,189}
\definecolor{steelblue31119180}{RGB}{31,119,180}

\begin{groupplot}[
  group style={
    group size=1 by 2,
    ylabels at=edge left,
    xlabels at=edge top,
    vertical sep=0.4cm,
    horizontal sep=0.1cm
  }]

\nextgroupplot[
legend cell align={left},
legend style={
  fill opacity=0.8,
  draw opacity=1,
  text opacity=1,
  at={(0.97,0.03)},
  anchor=south east,
  draw=lightgray204
},
tick align=outside,
ytick pos=left,
xtick pos=left,
width= \figwidthfull - 28.6pt,
height=\figheightfull/2,
x grid style={darkgray176},
xmajorgrids,
xmin=225, xmax=900,
xtick style={draw=none},
xtick={225, 450, 900},
xticklabels={},
y grid style={darkgray176},
ymajorgrids,
ymin=17.1, ymax=17.32,
ytick style={color=black}
]
\addplot [semithick, steelblue31119180, mark=*, mark size=3, mark options={solid}]
table {%
    225 17.19
    450 17.23
    900 17.28
};
\addlegendentry{t-DIP}

\addplot [semithick, darkorange25512714, mark=square*, mark size=3, mark options={solid}]
table {%
    225 17.16
    450 17.24
    900 17.3
};
\addlegendentry{FMLP}

\nextgroupplot[
legend cell align={left},
legend style={
  fill opacity=0.8,
  draw opacity=1,
  text opacity=1,
  at={(0.97,0.03)},
  anchor=south east,
  draw=lightgray204
},
tick align=outside,
tick pos=left,
width=\figwidthfull - 28.6pt,
height=\figheightfull/2,
x grid style={darkgray176},
xlabel={acquisition time in \SI{}{\second}},
xmajorgrids,
xmin=225, xmax=900,
xtick style={color=black},
xtick={225, 450, 900},
xticklabels={4, 8, 16},
y grid style={darkgray176},
ylabel style={align=center, xshift=0.85cm},
ylabel={SER within \\ \SI{4}{\second} of acquisition time},
ymajorgrids,
ymin=12.0, ymax=12.6,
ytick style={color=black}
]

\addplot [semithick, forestgreen4416044, mark=triangle*, mark size=3, mark options={solid}]
table {%
    225 12.53
    450 12.54
    900 12.54
};
\addlegendentry{KFMLP}

\addplot [semithick, mediumpurple148103189, mark=diamond*, mark size=3, mark options={solid}]
table {%
    225 12.08
    450 12.14
    900 12.14
};
\addlegendentry{NIK}

\end{groupplot}

\end{tikzpicture}
    \caption{
    The SER within the first \SI{4}{\second} of acquisition time improves with increasing the amount of measurement data beyond \SI{4}{\second}. The reconstruction quality can be improved by training the FMLP on a longer acquisition time. The methods were evaluated on the low-resolution high-SNR dataset with the same configurations as in Figure~\ref{fig:fmlp-tdip-bh-10-225-imgs} and $z_{\text{slack}} = 0.1, 0.2,$ and $0.4$ for $\frames=225$ (\SI{4}{\second}), $450$ (\SI{8}{\second}), and $900$ (\SI{16}{\second}), respectively.}
    \label{fig:fmlp-tdip-SER-225-450-900}
\end{figure}
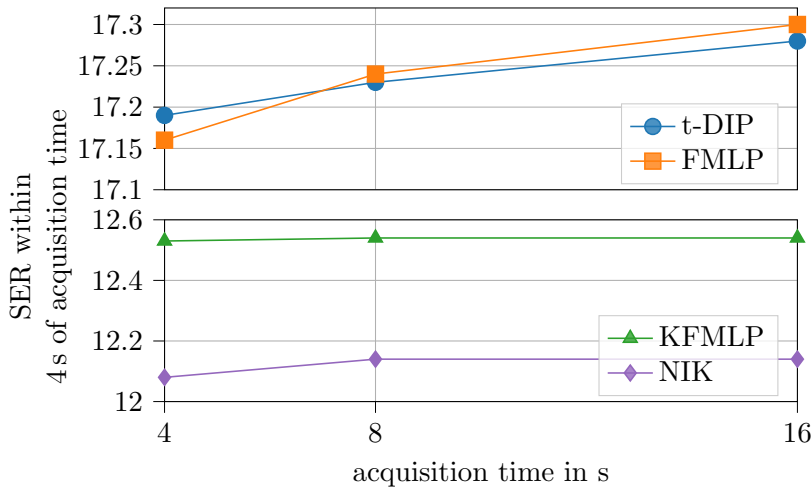

Figure~\ref{fig:fmlp-tdip-SER-225-450-900} shows that the SER of the FMLP and the t-DIP improves when training the methods on more data, indicating that both methods not only take advantage of temporal correlations between adjacent frames but also over the entire acquisition length. The NIK and the KFMLP, by contrast, do not improve substantially.

Remarkably, the FMLP and the t-DIP improve by a similar margin even though the t-DIP uses prior information about the heart rate that is encoded by the angular frequency of the helix trajectory. Information about the cardiac phase might be advantageous for correlating frames over multiple cardiac cycles. The FMLP does not incorporate such prior but nevertheless improves at a similar rate as a function of the acquisition length.
\vspace{1em}
\begin{table}[]
    \centering
    \caption{Maximum SER of the models on datasets in different operating regimes. FMLP and t-DIP perform similarly, while KFMLP performs significantly worse.}
    \begin{tabular}{c | c | c | c | c}
    \specialrule{.15em}{.05em}{.05em}
       dataset & FMLP & NIK & KFMLP & t-DIP \\
        \midrule
        low-resolution, high-SNR & 17.16 & 12.08 & 12.53 & \textbf{17.19} \\
        low-resolution, low-SNR  & \textbf{10.62} & 8.81 & 8.64 & 10.59 \\
        high-resolution & \textbf{9.00} & 6.90 & 6.51 & 8.97 \\
    \specialrule{.15em}{.05em}{.05em}
    \end{tabular}
    
    \label{tab:ser-datasets}
\end{table}

\subsection{Visual quality assessment}
\label{subsec:visual-assessment}

\begin{figure}[]
    \centering
    \input{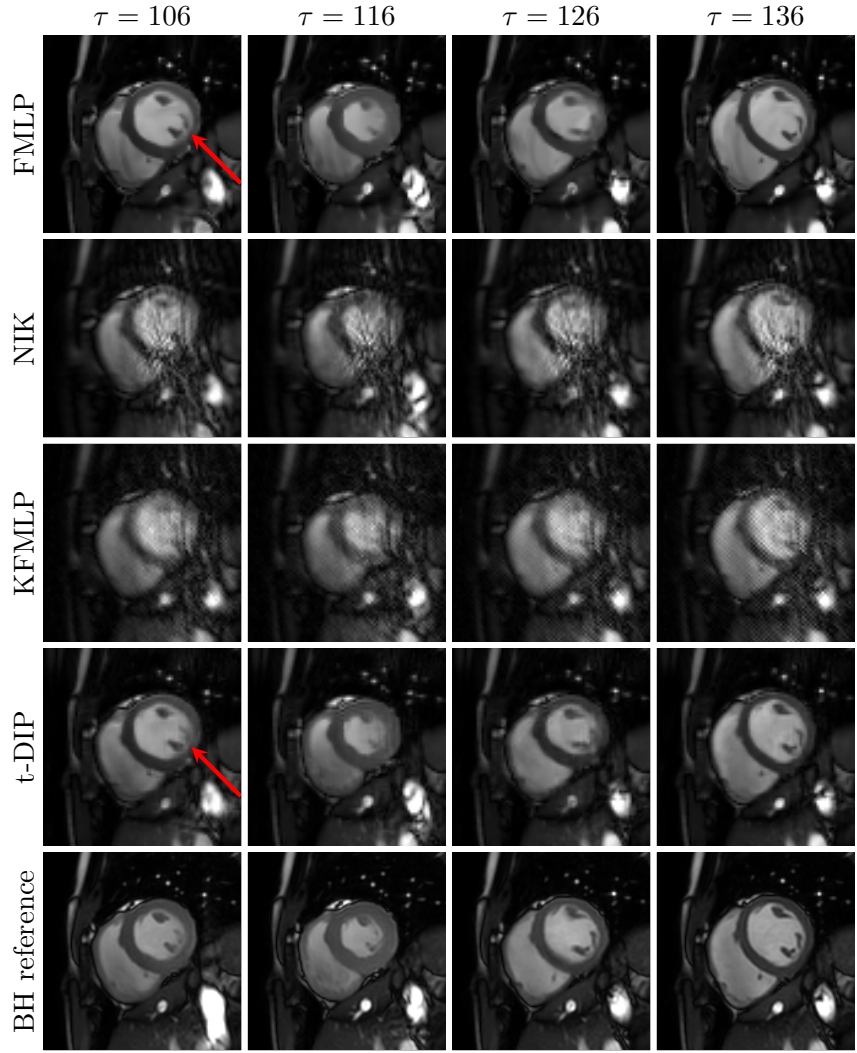}
    \vspace{1em}
    \caption{On the low-resolution high-SNR dataset, the image quality of the FMLP and the t-DIP are similar, whereby the FMLP recovers anatomic details such as the papillary muscles in the left ventricle (red arrow) more accurately. The reconstructions by the NIK and the KFMLP are distorted by aliasing-like artifacts and fine-structured noise such that anatomic details are not well recognizable.
    The models were trained on $\frames=225$ frames. 
    } 
    \label{fig:fmlp-tdip-bh-10-225-imgs}
\end{figure}

\begin{figure}[]
    \centering
    \input{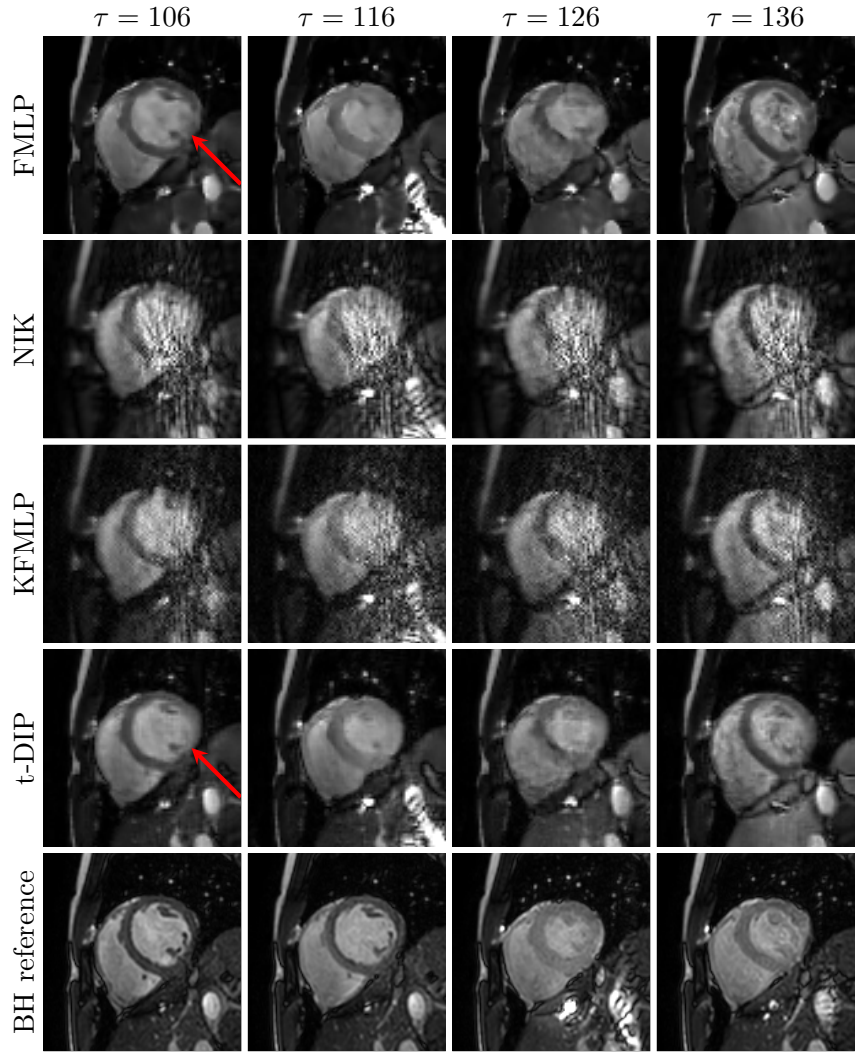}
    \vspace{1em}
    \caption{For an isotropic slice thickness (low-resolution low-SNR dataset) the reconstruction quality of all methods degrades and new artifacts are introduced. The FMLP and the t-DIP achieve a similar image quality and outperform the NIK and the KFMLP. The models were trained on $\frames=225$ frames. 
    }
    \label{fig:fmlp-tdip-bh-15-225-imgs}
\end{figure}

\begin{figure}[ht] 
    \centering
    \input{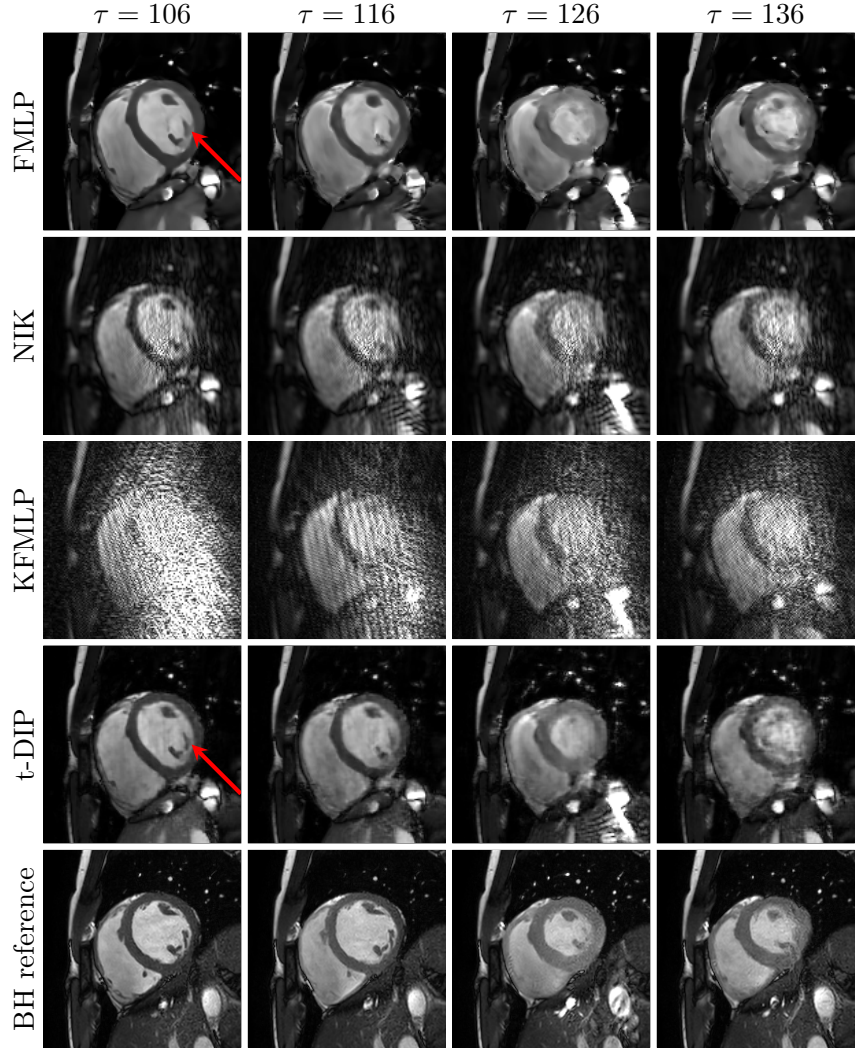}
    \vspace{1em}
    \caption{On the high-resolution dataset, the FMLP and the t-DIP achieve a similar reconstruction quality and both methods suffer from similar artifacts. The quality of the NIK and the KFMLP is decreased substantially by noise-like artifacts, especially for the KFMLP. The models were trained on $\frames = 225$ frames. 
    }
    \label{fig:fmlp-tdip-bh-20-225-imgs}
\end{figure}

We compare the reconstructed images of the FMLP, NIK, KFMLP, and t-DIP to the ECG-triggered breath-hold reconstructions that serve as a visual reference. The methods are compared on the low-resolution high-SNR, the low-resolution low-SNR, and the high-resolution dataset. For each dataset, we show reconstructions of four frames that have been selected to cover a cardiac cycle.

The reconstructions in the Figures~\ref{fig:fmlp-tdip-bh-10-225-imgs}, \ref{fig:fmlp-tdip-bh-15-225-imgs}, and \ref{fig:fmlp-tdip-bh-20-225-imgs}, show that FMLP performs on par with the t-DIP in terms of perceived image quality. Although the overall differences in quality are marginal, the FMLP recovers anatomic details slightly more accurately on the low-resolution high-SNR dataset. The differences are most apparent in small moving anatomic structures of the heart, such as the papillary muscles, see Figure~\ref{fig:fmlp-tdip-bh-10-225-imgs}.
On the low-resolution low-SNR and the high-resolution datasets, the FMLP and the t-DIP exhibit similar artifacts, see Figures~\ref{fig:fmlp-tdip-bh-15-225-imgs}, and \ref{fig:fmlp-tdip-bh-20-225-imgs}.

We find that the FMLP achieves a better image quality on all of the datasets when compared to the NIK and the KFMLP. On the low-resolution high-SNR dataset, the NIK and the KFMLP suffer aliasing-like artifacts that blur anatomic details, such as the papillary muscles, see Figure~\ref{fig:fmlp-tdip-bh-10-225-imgs}. On the low-resolution low-SNR dataset, the aliasing-like artifacts are superimposed by the noise that further degrades the image quality compared to the FMLP (Figure~\ref{fig:fmlp-tdip-bh-15-225-imgs}).

On the high-resolution dataset, the noise becomes most prominent for the KFMLP, where it obscures any meaningful details in the cardiac region. The NIK performs better than the KFMLP, but still suffers from aliasing-like artifacts and noise unlike the FMLP that reconstructs more clean and sharp images with a higher fidelity of details (Figure~\ref{fig:fmlp-tdip-bh-20-225-imgs}).

We note that the presented reconstructions by the NIK are worse than the reconstructions shown by the authors Huang et al. \cite{huangNeuralImplicitKSpace2022}. However, the NIK has been tested on a radial tiny golden-angle sampling pattern that eventually covers the entire k-space, hence providing information on the entire k-space. For our datasets, by contrast, the majority of k-space lines are never sampled, thus, requiring a  strong prior of the untrained network. Moreover, the datasets reconstructed by the NIK were acquired during breath-holding, which eliminates artifacts due to respiratory motion. 


\begin{table*}[ht]
    \centering
    \caption{Computational performance characteristics.}
    \begin{tabular}{c | c | c | c | c | c}
    \specialrule{.15em}{.05em}{.05em}
        res. & property & FMLP & NIK & KFMLP & t-DIP \\
        \midrule
        \multirow{5}{*}{\rotatebox[origin=c]{90}{$186 \times 264$}} & \# parameters & \num{1.91e6} & \num{2.10e6}& \num{1.93e6} & \num{5.94e6} \\
                         & memory on GPU &  \SI{3.55}{\giga\byte} & \SI{2.26}{\giga\byte} & \SI{1.14}{\giga\byte} & \SI{1.94}{\giga\byte} \\
                         & time per epoch & \SI{19.92}{\second} & \SI{23.96}{\second} & \SI{4.06}{\second} & \SI{8.56}{\second} \\
                         & number of epochs & 744 & 1040 & 309 & 252 \\
                         & total training time & \SI{510}{\minute} & \SI{624}{\minute} & \SI{35}{\minute} & \SI{42}{\minute}\\
        \midrule
        \multirow{5}{*}{\rotatebox[origin=c]{90}{$334 \times 480$}} & \# parameters & \num{1.91e6} & \num{2.10e6} & \num{1.93e6} & \num{7.12e6} \\
                         & memory on GPU & \SI{9.21}{\giga\byte} & \SI{2.33}{\giga\byte} & \SI{1.22}{\giga\byte} & \SI{4.03}{\giga\byte} \\
                         & time per epoch & \SI{63.82}{\second} & \SI{34.66}{\second} & \SI{4.11}{\second} & \SI{26.65}{\second} \\
                         & number of epochs & 1872 & 930 & 601 & 176 \\
                         & total training time & \SI{3237}{\minute} & \SI{780}{\minute}& \SI{110}{\minute} & \SI{91}{\minute}\\
    \specialrule{.15em}{.05em}{.05em}
    \end{tabular}
    \label{tab:computation}
\end{table*}

\subsection{Computational cost}
\label{subsec:computational-demands}

The computational costs depend on the hyper-parameter configuration, but for most of our tested hyper-parameter configurations, training the FMLP requires more epochs and takes longer to train per epoch than the t-DIP and the KFMLP. 
The KFMLP and the t-DIP reach the maximum SER in a similar amount of training time, whereas the t-DIP converges in fewer epochs. The NIK takes a longer time to train per epoch than the KFMLP and the t-DIP and requires more epochs, amounting in a longer training time. 

The computational performance characteristics of the configurations with the highest SER are listed in Table~\ref{tab:computation}. 
The total training time is measured by the wall clock time. 
The time per epoch is measured by the wall clock time over 100 epochs of training without additional overhead. The hyper-parameter configurations with maximum SER are chosen that are specified in Figure~\ref{fig:fmlp-tdip-bh-10-225-imgs} and \ref{fig:fmlp-tdip-bh-20-225-imgs}.  

The computational cost scales differently with image resolution for the three methods.
An advantage of the FMLP is that the network size can stay constant as the image resolution is increased. However, the network has to be evaluated for all pixels and, thus, the computational cost scales linearly with image resolution. The CNN of the t-DIP, by contrast, needs to be adapted to an increasing image resolution by adding up-sampling layers and convolutional layers. Thus, the number of trainable parameters increases with image resolution. However, doubling the image resolution merely requires adding one more stage of up-sampling and convolutional layers. Thus, the cost of evaluating the CNN scales efficiently with image resolution. The network size of the NIK and the KFMLP can also be set independently of the image resolution. 

Moreover, the computational cost of the NIK and the KFMLP is low compared to the FMLP, even though all three use a neural network of similar size. However, a training step of the FMLP requires an evaluation of the network on all coordinates on the grid, i.e., on all pixels of the image, whereas the NIK and the KFMLP are only evaluated on the measured coordinates. Since each frame is largely undersampled, the number of evaluations is much lower for the NIK and KFMLP than for the FMLP. The NIK and the KFMLP are evaluated on the full Cartesian grid only once after training. 
 

\section{Discussion and Conclusion}
\label{sec:conclusion}

We proposed an untrained reconstruction method based on implicit networks, called FMLP, for cardiac real-time MRI. The method uses a ReLU-MLP with Fourier-feature inputs to encode spatio-temporal coordinate vectors for representing an image of the beating heart. We evaluated the method on experimental datasets for 2D free-breathing cardiac real-time MRI covering different operating regimes, i.e., different image resolutions, slice thicknesses, and acquisition lengths. 

We find that FMLP can achieve reconstruction quality on par or slightly better with the best CNN-based untrained methods (t-DIP), as measured with quantitative metrics (SER) and as determined through visual comparisons. 
FMLP tends to improve faster as a function of the acquisition time (see Figure~\ref{fig:fmlp-tdip-SER-225-450-900}), and slightly surpassed t-DIP for longer acquisition times. 

We note that the similarity in performance to the t-DIP is somewhat expected since MLPs with Fourier-feature inputs and CNNs both have a bias towards fitting 
smooth images and low-frequency contents of the signal first~\cite{tancikFourierFeaturesLet2020, heckelDenoisingRegularizationExploiting2020}. 
This implicit bias suppresses high-frequency aliasing artifacts, thereby interpolating the missing frequencies in the k-space and reconstructing the image.

While an MLP with Fourier-features is a good image model, it is not a good model for representing an image in the Fourier domain. We found that the NIK and the KFMLP, which rely on representing the signal in the Fourier domain with a Fourier-feature MLP, suffered from aliasing-like and noise-like artifacts that distorted the images considerably. These artifacts arise since a Fourier-feature MLP applied in the Fourier domain does not interpolate images well. 
Although the NIK and KFMLP do a good job of representing the measurements along the sampling trajectory, they do not predict frequencies that had not been well captured by the measurement. 

We note that the partial-Fourier random Cartesian sampling pattern used for the experiments may pose an especially difficult challenge to the reconstruction methods as a small subset of k-space lines is measured repetitively. Thus, a large fraction of the k-space is not measured at any time and needs to be interpolated with an efficient image model. 

Visually, the KFMLP and the NIK achieve a similar image quality on the low-resolution high-SNR dataset and the low-resolution low-SNR dataset, which is expected since both methods are similar and fit data directly in the k-space. However, the KFMLP exhibits substantially more noise on the high-resolution dataset than the NIK. 

We hasten to add that the image quality of our FMLP method comes at a high computational cost. The FMLP is more expensive to fit than the t-DIP and the KFMLP.
However, we think that future research (such as pre-training the FMLP) can substantially improve the running time of the FMLP, and different feature encoding altogether might also improve the running time. 

Compared to the FMLP, the NIK demands similar computational resources. Its training is slow due to the evaluation of the k-space for each receiver coil. For instance, our datasets comprised 25 receiver coils in total. The KFMLP, improves on that by evaluating all receiver coils in one forward pass.

A major advantage of the FMLP over the t-DIP and other approaches is its flexibility: To generalize to 3D data, we only need to add an additional spatial input. Thus, the the FMLP, as well as other coordinate-based approaches such as the NIK and the KFMLP can easily be extended to 3D real-time data.


%



\section*{Acknowledgements}

The authors would like to thank you Kilian Weiss (Philips GmbH Market DACH, Hamburg, Germany) for helpful discussion. 
The work is supported by the Deutsche Forschungsgemeinschaft (DFG, German Research Foundation)
11 - 456465471, 464123524.

\section*{References}
\printbibliography[heading=none]

\onecolumn

\clearpage

\begin{appendices}

\section{Ablation Studies of the FMLP}

In this section, we conduct ablation studies on the hyperparameters of the FMLP, specifically on the number of hidden layers and neurons per layer of the MLP, and on the spatial and temporal coordinate scales  which are important parameters for controlling the spatial and temporal prior imposed on the heart to be reconstructed. 
We perform grid-searches over those parameters and study the impact on the reconstruction quality and computational costs. 

We also study whether the reconstruction quality can be improved by adding an additional loss term that explicitly regularizes the temporal total variation. We find that if the temporal coordinate scale is tuned properly, additional temporal total variation regularization is not beneficial. 

The network architecture of the FMLP is critical for performance. We compare two variants of the FMLP with joint and separate Fourier-feature embeddings for the spatial and temporal coordinates. We find that separate embeddings yield an improved reconstruction quality.

All experiments in this section are conducted on the low-resolution high-SNR dataset with $\frames = 225$ frames corresponding to $\SI{4}{\second}$ acquisition length. We train the FMLP until no new SER highscore has been reached in the last 200 epochs and report the performance at the epochs of maximum SER.

For the FMLP, we jointly tune the spatial coordinates $\csx$ and $\csy$ with a fixed ratio. Initially, we intended setting $\csx = \csy$, however, we the field-of-view (FOV) of the experimental datasets is enlarged due to over-sampling along the frequency encoding ($\ky$) and phase encoding ($\ky$) dimensions. Thus, we calculated the coordinate grid for a smaller FOV than was actually measured, resulting in new effective coordinate scales with $\csy = 1.43\csx$ in the ablation studies. We find, however, that the skewed coordinate scales yield very similar results to setting $\csy = \csx$.

For simplicity, we define a default hyperparameter configuration for the ablation studies. Unless not stated explicitly, the parameters are set to: spatial coordinate scales $\csy = 1.43\csx = \SI{33}{\metre^{-1}}$, temporal coordinate scale $\cst = \SI{5.3}{\second^{-1}}$, number of lines per frame $\numlines = 6$, number of hidden layers $\hiddenlayers = 7$, and 512 neurons per hidden layer.

\subsection{Spatio-temporal coordinate scales regularize the smoothness along spatio-temporal dimensions}

The spatio-temporal coordinate scales of the FMLP effectively adjust the regularization strength in the spatial and temporal dimensions. We perform grid-searches over the spatial coordinate scales $\csx, \csy$ and the temporal coordinate scale $\cst$. For all configurations, we report the maximum SER and the number of training epochs required to reach the maximum SER.

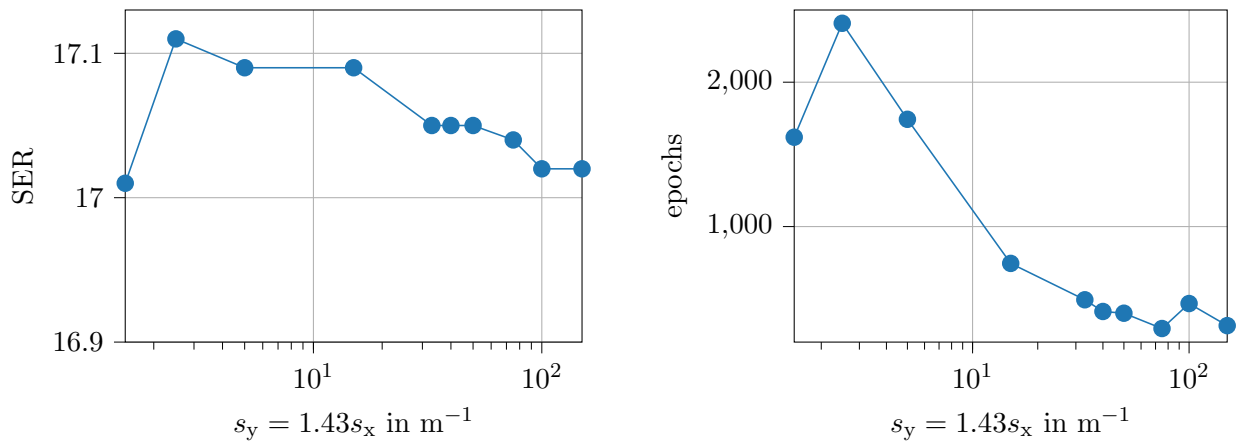
\begin{figure}[h]
    \begin{subfigure}[t]{0.47\textwidth}
\begin{tikzpicture}

\definecolor{crimson2143940}{RGB}{214,39,40}
\definecolor{darkgray176}{RGB}{176,176,176}
\definecolor{darkorange25512714}{RGB}{255,127,14}
\definecolor{forestgreen4416044}{RGB}{44,160,44}
\definecolor{lightgray204}{RGB}{204,204,204}
\definecolor{mediumpurple148103189}{RGB}{148,103,189}
\definecolor{steelblue31119180}{RGB}{31,119,180}

\begin{semilogxaxis}[
legend cell align={left},
legend style={
  fill opacity=0.8,
  draw opacity=1,
  text opacity=1,
  at={(0.97,0.97)},
  anchor=north east,
  draw=lightgray204
},
tick align=outside,
tick pos=left,
width=\figwidthhalf - 3pt,
height=6cm,
x grid style={darkgray176},
xlabel={$\csy = 1.43\csx$ in \SI{}{\metre^{-1}}},
xmajorgrids,
xmin=1.5, xmax=150,
xtick style={color=black},
y grid style={darkgray176},
ylabel={SER},
ymajorgrids,
ymin=16.9, ymax=17.13,
ytick style={color=black}
]
\addplot [semithick, steelblue31119180, mark=*, mark size=3, mark options={solid}]
table {%
    1.5 17.01
    2.5 17.11
    5 17.09
    15 17.09
    33 17.05
    40 17.05
    50 17.05
    75 17.04
    100 17.02
    150 17.02
};



%


\end{semilogxaxis}

\end{tikzpicture}
        \caption{The image quality (in SER), as a function of the spacial coordinate $\csy$, which is coupled to the scale $\csx$ as  $\csy=1.43\csx$. 
        It can be observed that the SER increases as the spatial coordinate scales decrease, until the SER drops for very low coordinate scales, e.g., $\csy = 1.43\csy = \SI{1.5}{\metre^{-1}}$.}
        \label{fig:spatial-coordinate-scales-SER}
    \end{subfigure}
    \hfill
    \begin{subfigure}[t]{0.47\textwidth}
\begin{tikzpicture}

\definecolor{crimson2143940}{RGB}{214,39,40}
\definecolor{darkgray176}{RGB}{176,176,176}
\definecolor{darkorange25512714}{RGB}{255,127,14}
\definecolor{forestgreen4416044}{RGB}{44,160,44}
\definecolor{lightgray204}{RGB}{204,204,204}
\definecolor{mediumpurple148103189}{RGB}{148,103,189}
\definecolor{steelblue31119180}{RGB}{31,119,180}

\begin{semilogxaxis}[
legend cell align={left},
legend style={
  fill opacity=0.8,
  draw opacity=1,
  text opacity=1,
  at={(0.97,0.97)},
  anchor=north east,
  draw=lightgray204
},
tick align=outside,
tick pos=left,
width=\figwidthhalf - 12pt,
height=6cm,
x grid style={darkgray176},
xlabel={$\csy = 1.43\csx$ in \SI{}{\metre^{-1}}},
xmajorgrids,
xmin=1.5, xmax=150,
xtick style={color=black},
y grid style={darkgray176},
ylabel={epochs},
ymajorgrids,
ymin=200, ymax=2500,
ytick style={color=black}
]
\addplot [semithick, steelblue31119180, mark=*, mark size=3, mark options={solid}]
table {%
    1.5 1619
    2.5 2408
    5 1743
    15 745
    33 493
    40 412
    50 400
    75 294
    100 467
    150 314
};


\end{semilogxaxis}

\end{tikzpicture}
        \caption{The number of training epochs as a function of the spatial coordinate scales $\csy=1.43\csx$. The number of epochs increases for decreasing spatial coordinate scales until a threshold. Below the threshold, the number of epochs drops. The drop is may be caused by the slow convergence and the stopping criterion that stops training after 200 epochs without new SER highscore.}
        \label{fig:spatial-coordinate-scales-epochs}
    \end{subfigure}
    \vspace{0.5em}
    \caption{Choosing the spatial coordinate scales $\csx, \csy$ is a trade-off between SER and training time.}
\end{figure}

\textbf{Performance as a function of spatial coordinate scales: }
Figure~\ref{fig:spatial-coordinate-scales-SER} shows that the image quality (in SER) as a function of the spatial coordinate scales first improves and then deteriorates, as expected. Too small and too large values result in too little and too much regularization.

Figure~\ref{fig:spatial-coordinate-scales-epochs} shows the number of training epochs required for reaching the maximum SER for different spatial coordinate scales. 
The figure shows that a smaller number of epochs is required to reach the best SER for a given scale as a function of the spatial coordinate scales. 
However, for coordinate scales below the threshold of $\csy = 1.43\csx = \SI{2.5}{\metre^{-1}}$, the number of epochs suddenly decreases. The peculiar decrease may be attributed to the termination policy that stops training once no SER highscore has been achieved within the last 200 epochs. For very low coordinate scales, the convergence is so slow that only small improvements are achieved within 200 epochs. Small jitters in the SER score cause the training to stop early.

\begin{figure}[h]
    \centering
\begin{tikzpicture}

\newcommand\imgwidth[0]{4.8cm}

\definecolor{darkgray176}{RGB}{176,176,176}

\begin{groupplot}[
  group style={
    group size=4 by 4,
    ylabels at=edge left,
    xlabels at=edge top,
    vertical sep=0.1cm,
    horizontal sep=0.1cm
  }]

\nextgroupplot[
scaled x ticks=manual:{}{\pgfmathparse{#1}},
scaled y ticks=manual:{}{\pgfmathparse{#1}},
width=\imgwidth,
height=\imgwidth,
tick align=outside,
x grid style={darkgray176},
xmajorticks=false,
xmin=0, xmax=200,
xtick style={color=black},
xticklabels={},
y dir=reverse,
y grid style={darkgray176},
ymajorticks=false,
ymin=0, ymax=200,
ytick style={color=black},
yticklabels={},
ylabel={$\frameindex = 106$},
xlabel={$\csy = \SI{33}{\metre^{-1}}$}
]
\addplot graphics [includegraphics cmd=\pgfimage, xmin=0, xmax=200, ymin=200, ymax=0] {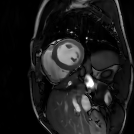};

\nextgroupplot[
scaled x ticks=manual:{}{\pgfmathparse{#1}},
scaled y ticks=manual:{}{\pgfmathparse{#1}},
width=\imgwidth,
height=\imgwidth,
tick align=outside,
x grid style={darkgray176},
xmajorticks=false,
xmin=0, xmax=200,
xtick style={color=black},
xticklabels={},
y dir=reverse,
y grid style={darkgray176},
ymajorticks=false,
ymin=0, ymax=200,
ytick style={color=black},
yticklabels={},
xlabel={$\csy = \SI{75}{\metre^{-1}}$}
]
\addplot graphics [includegraphics cmd=\pgfimage, xmin=0, xmax=200, ymin=200, ymax=0] {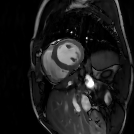};

\nextgroupplot[
scaled x ticks=manual:{}{\pgfmathparse{#1}},
scaled y ticks=manual:{}{\pgfmathparse{#1}},
width=\imgwidth,
height=\imgwidth,
tick align=outside,
x grid style={darkgray176},
xmajorticks=false,
xmin=0, xmax=200,
xtick style={color=black},
xticklabels={},
y dir=reverse,
y grid style={darkgray176},
ymajorticks=false,
ymin=0, ymax=200,
ytick style={color=black},
yticklabels={},
xlabel={$\csy = \SI{150}{\metre^{-1}}$}
]
\addplot graphics [includegraphics cmd=\pgfimage, xmin=0, xmax=200, ymin=200, ymax=0] {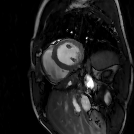};

\nextgroupplot[
scaled x ticks=manual:{}{\pgfmathparse{#1}},
scaled y ticks=manual:{}{\pgfmathparse{#1}},
width=\imgwidth,
height=\imgwidth,
tick align=outside,
x grid style={darkgray176},
xmajorticks=false,
xmin=0, xmax=200,
xtick style={color=black},
xticklabels={},
y dir=reverse,
y grid style={darkgray176},
ymajorticks=false,
ymin=0, ymax=200,
ytick style={color=black},
yticklabels={},
xlabel={BH reference}
]
\addplot graphics [includegraphics cmd=\pgfimage, xmin=0, xmax=200, ymin=200, ymax=0] {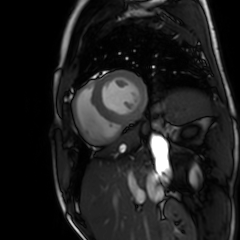};

\nextgroupplot[
scaled x ticks=manual:{}{\pgfmathparse{#1}},
scaled y ticks=manual:{}{\pgfmathparse{#1}},
width=\imgwidth,
height=\imgwidth,
tick align=outside,
x grid style={darkgray176},
xmajorticks=false,
xmin=0, xmax=200,
xtick style={color=black},
xticklabels={},
y dir=reverse,
y grid style={darkgray176},
ymajorticks=false,
ymin=0, ymax=200,
ytick style={color=black},
yticklabels={},
ylabel={$\frameindex = 116$},
]
\addplot graphics [includegraphics cmd=\pgfimage, xmin=0, xmax=200, ymin=200, ymax=0] {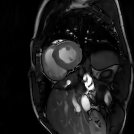};

\nextgroupplot[
scaled x ticks=manual:{}{\pgfmathparse{#1}},
scaled y ticks=manual:{}{\pgfmathparse{#1}},
width=\imgwidth,
height=\imgwidth,
tick align=outside,
x grid style={darkgray176},
xmajorticks=false,
xmin=0, xmax=200,
xtick style={color=black},
xticklabels={},
y dir=reverse,
y grid style={darkgray176},
ymajorticks=false,
ymin=0, ymax=200,
ytick style={color=black},
yticklabels={}
]
\addplot graphics [includegraphics cmd=\pgfimage, xmin=0, xmax=200, ymin=200, ymax=0] {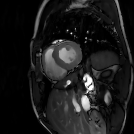};

\nextgroupplot[
scaled x ticks=manual:{}{\pgfmathparse{#1}},
scaled y ticks=manual:{}{\pgfmathparse{#1}},
width=\imgwidth,
height=\imgwidth,
tick align=outside,
x grid style={darkgray176},
xmajorticks=false,
xmin=0, xmax=200,
xtick style={color=black},
xticklabels={},
y dir=reverse,
y grid style={darkgray176},
ymajorticks=false,
ymin=0, ymax=200,
ytick style={color=black},
yticklabels={}
]
\addplot graphics [includegraphics cmd=\pgfimage, xmin=0, xmax=200, ymin=200, ymax=0] {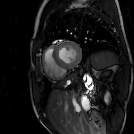};

\nextgroupplot[
scaled x ticks=manual:{}{\pgfmathparse{#1}},
scaled y ticks=manual:{}{\pgfmathparse{#1}},
width=\imgwidth,
height=\imgwidth,
tick align=outside,
x grid style={darkgray176},
xmajorticks=false,
xmin=0, xmax=200,
xtick style={color=black},
xticklabels={},
y dir=reverse,
y grid style={darkgray176},
ymajorticks=false,
ymin=0, ymax=200,
ytick style={color=black},
yticklabels={}
]
\addplot graphics [includegraphics cmd=\pgfimage, xmin=0, xmax=200, ymin=200, ymax=0] {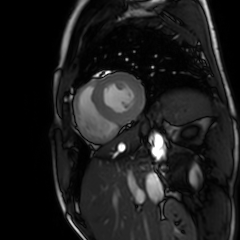};

\nextgroupplot[
scaled x ticks=manual:{}{\pgfmathparse{#1}},
scaled y ticks=manual:{}{\pgfmathparse{#1}},
width=\imgwidth,
height=\imgwidth,
tick align=outside,
x grid style={darkgray176},
xmajorticks=false,
xmin=0, xmax=200,
xtick style={color=black},
xticklabels={},
y dir=reverse,
y grid style={darkgray176},
ymajorticks=false,
ymin=0, ymax=200,
ytick style={color=black},
yticklabels={},
ylabel={$\frameindex = 126$},
]
\addplot graphics [includegraphics cmd=\pgfimage, xmin=0, xmax=200, ymin=200, ymax=0] {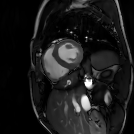};

\nextgroupplot[
scaled x ticks=manual:{}{\pgfmathparse{#1}},
scaled y ticks=manual:{}{\pgfmathparse{#1}},
width=\imgwidth,
height=\imgwidth,
tick align=outside,
x grid style={darkgray176},
xmajorticks=false,
xmin=0, xmax=200,
xtick style={color=black},
xticklabels={},
y dir=reverse,
y grid style={darkgray176},
ymajorticks=false,
ymin=0, ymax=200,
ytick style={color=black},
yticklabels={}
]
\addplot graphics [includegraphics cmd=\pgfimage, xmin=0, xmax=200, ymin=200, ymax=0] {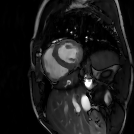};

\nextgroupplot[
scaled x ticks=manual:{}{\pgfmathparse{#1}},
scaled y ticks=manual:{}{\pgfmathparse{#1}},
width=\imgwidth,
height=\imgwidth,
tick align=outside,
x grid style={darkgray176},
xmajorticks=false,
xmin=0, xmax=200,
xtick style={color=black},
xticklabels={},
y dir=reverse,
y grid style={darkgray176},
ymajorticks=false,
ymin=0, ymax=200,
ytick style={color=black},
yticklabels={}
]
\addplot graphics [includegraphics cmd=\pgfimage, xmin=0, xmax=200, ymin=200, ymax=0] {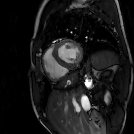};

\nextgroupplot[
scaled x ticks=manual:{}{\pgfmathparse{#1}},
scaled y ticks=manual:{}{\pgfmathparse{#1}},
width=\imgwidth,
height=\imgwidth,
tick align=outside,
x grid style={darkgray176},
xmajorticks=false,
xmin=0, xmax=200,
xtick style={color=black},
xticklabels={},
y dir=reverse,
y grid style={darkgray176},
ymajorticks=false,
ymin=0, ymax=200,
ytick style={color=black},
yticklabels={}
]
\addplot graphics [includegraphics cmd=\pgfimage, xmin=0, xmax=200, ymin=200, ymax=0] {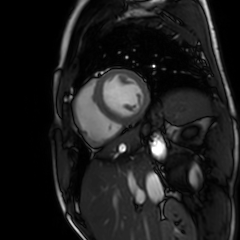};

\nextgroupplot[
scaled x ticks=manual:{}{\pgfmathparse{#1}},
scaled y ticks=manual:{}{\pgfmathparse{#1}},
width=\imgwidth,
height=\imgwidth,
tick align=outside,
x grid style={darkgray176},
xmajorticks=false,
xmin=0, xmax=200,
xtick style={color=black},
xticklabels={},
y dir=reverse,
y grid style={darkgray176},
ymajorticks=false,
ymin=0, ymax=200,
ytick style={color=black},
yticklabels={},
ylabel={$\frameindex = 136$},
]
\addplot graphics [includegraphics cmd=\pgfimage, xmin=0, xmax=200, ymin=200, ymax=0] {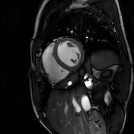};

\nextgroupplot[
scaled x ticks=manual:{}{\pgfmathparse{#1}},
scaled y ticks=manual:{}{\pgfmathparse{#1}},
width=\imgwidth,
height=\imgwidth,
tick align=outside,
x grid style={darkgray176},
xmajorticks=false,
xmin=0, xmax=200,
xtick style={color=black},
xticklabels={},
y dir=reverse,
y grid style={darkgray176},
ymajorticks=false,
ymin=0, ymax=200,
ytick style={color=black},
yticklabels={}
]
\addplot graphics [includegraphics cmd=\pgfimage, xmin=0, xmax=200, ymin=200, ymax=0] {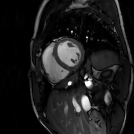};

\nextgroupplot[
scaled x ticks=manual:{}{\pgfmathparse{#1}},
scaled y ticks=manual:{}{\pgfmathparse{#1}},
width=\imgwidth,
height=\imgwidth,
tick align=outside,
x grid style={darkgray176},
xmajorticks=false,
xmin=0, xmax=200,
xtick style={color=black},
xticklabels={},
y dir=reverse,
y grid style={darkgray176},
ymajorticks=false,
ymin=0, ymax=200,
ytick style={color=black},
yticklabels={}
]
\addplot graphics [includegraphics cmd=\pgfimage, xmin=0, xmax=200, ymin=200, ymax=0] {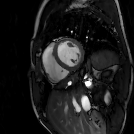};

\nextgroupplot[
scaled x ticks=manual:{}{\pgfmathparse{#1}},
scaled y ticks=manual:{}{\pgfmathparse{#1}},
width=\imgwidth,
height=\imgwidth,
tick align=outside,
x grid style={darkgray176},
xmajorticks=false,
xmin=0, xmax=200,
xtick style={color=black},
xticklabels={},
y dir=reverse,
y grid style={darkgray176},
ymajorticks=false,
ymin=0, ymax=200,
ytick style={color=black},
yticklabels={}
]
\addplot graphics [includegraphics cmd=\pgfimage, xmin=0, xmax=200, ymin=200, ymax=0] {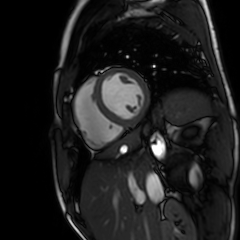};

\end{groupplot}

\end{tikzpicture}
    \vspace{1em}
    \caption{Reconstructions are shown for different spatial coordinate scales. The smaller the coordinate scales, the smoother the image is. The spatial coordinate scales $\csx$ and $\csy$ control the regularization along the spatial dimensions.}
    \label{fig:spatial-coordinate-scales-imgs}
\end{figure}

Figure~\ref{fig:spatial-coordinate-scales-imgs} shows images reconstructed at different spatial coordinate scales. It can be seen that lower coordinate scales yield slightly smoother images. The spatial coordinate scales provide control over the smoothness of the reconstructions.

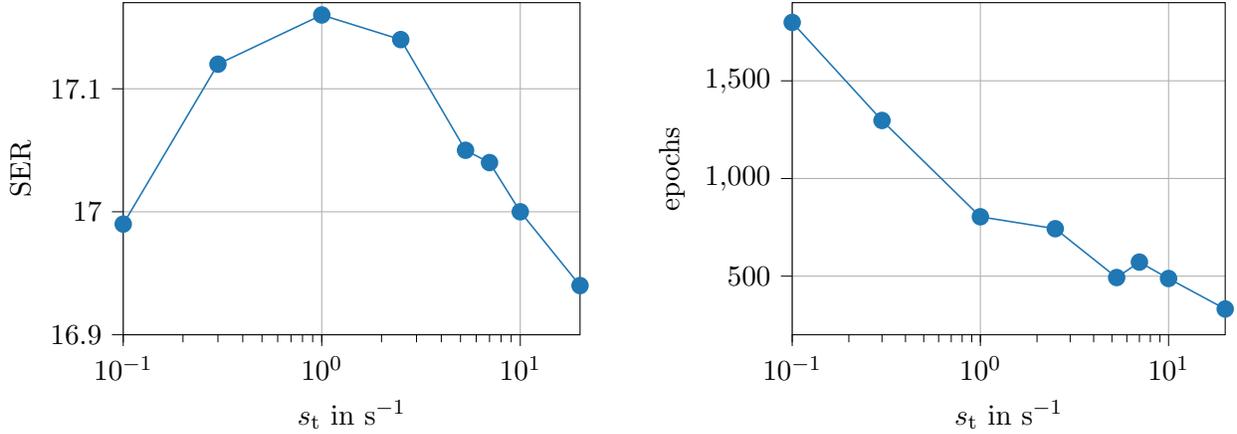
\begin{figure}[h]
    \centering
    \begin{subfigure}[t]{0.47\textwidth}
\begin{tikzpicture}

\definecolor{crimson2143940}{RGB}{214,39,40}
\definecolor{darkgray176}{RGB}{176,176,176}
\definecolor{darkorange25512714}{RGB}{255,127,14}
\definecolor{forestgreen4416044}{RGB}{44,160,44}
\definecolor{lightgray204}{RGB}{204,204,204}
\definecolor{mediumpurple148103189}{RGB}{148,103,189}
\definecolor{steelblue31119180}{RGB}{31,119,180}

\begin{semilogxaxis}[
legend cell align={left},
legend style={
  fill opacity=0.8,
  draw opacity=1,
  text opacity=1,
  at={(0.97,0.97)},
  anchor=north east,
  draw=lightgray204
},
tick align=outside,
tick pos=left,
width=\figwidthhalf-3pt,
height=6cm,
x grid style={darkgray176},
xlabel={$\cst$ in \SI{}{\second^{-1}}},
xmajorgrids,
xmin=0.1, xmax=20,
xtick style={color=black},
y grid style={darkgray176},
ylabel={SER},
ymajorgrids,
ymin=16.9, ymax=17.17,
ytick style={color=black}
]
\addplot [semithick, steelblue31119180, mark=*, mark size=3, mark options={solid}]
table {%
    0.1 16.99
    0.3 17.12
    1.0 17.16
    2.5 17.14
    5.3 17.05
    7 17.04
    10 17.00
    20 16.94
};

\end{semilogxaxis}

\end{tikzpicture}
        \caption{The image quality measured by the SER is plotted over a range of temporal coordinate scales. Among the tested configurations, the SER reaches its maximum for $\cst = \SI{1}{\second^{-1}}$ and drops for lower and higher scales.}
        \label{fig:temporal-coordinate-scales-SER}
    \end{subfigure}
    \hfill
    \begin{subfigure}[t]{0.47\textwidth}
\begin{tikzpicture}

\definecolor{crimson2143940}{RGB}{214,39,40}
\definecolor{darkgray176}{RGB}{176,176,176}
\definecolor{darkorange25512714}{RGB}{255,127,14}
\definecolor{forestgreen4416044}{RGB}{44,160,44}
\definecolor{lightgray204}{RGB}{204,204,204}
\definecolor{mediumpurple148103189}{RGB}{148,103,189}
\definecolor{steelblue31119180}{RGB}{31,119,180}

\begin{semilogxaxis}[
legend cell align={left},
legend style={
  fill opacity=0.8,
  draw opacity=1,
  text opacity=1,
  at={(0.97,0.97)},
  anchor=north east,
  draw=lightgray204
},
tick align=outside,
tick pos=left,
width=\figwidthhalf-12pt,
height=6cm,
x grid style={darkgray176},
xlabel={$\cst$ in \SI{}{\second^{-1}}},
xmajorgrids,
xmin=0.1, xmax=20,
xtick style={color=black},
y grid style={darkgray176},
ylabel={epochs},
ymajorgrids,
ymin=200, ymax=1900,
ytick style={color=black}
]
\addplot [semithick, steelblue31119180, mark=*, mark size=3, mark options={solid}]
table {%
    0.1 1799
    0.3 1297
    1.0 804
    2.5 743
    5.3 493
    7 572
    10 488
    20 332
};


\end{semilogxaxis}

\end{tikzpicture}
        \caption{The number of epochs is shown a function of the spatial coordinate scale. It increases as the temporal coordinate scale is decreased.}
        \label{fig:temporal-coordinate-scales-epochs}
    \end{subfigure}
    \vspace{0.5em}
    \caption{The temporal coordinate scale $\cst$ affects the maximum SER and the training time.}
\end{figure}

\textbf{Performance as a function of the temporal coordinate scale: }
In Figure~\ref{fig:temporal-coordinate-scales-SER}, the image quality (in SER) is plotted over a range of temporal coordinate scales. The SER reaches its maximum for $\cst = \SI{1}{\second^{-1}}$ and gradually decreases for lower and higher spatial coordinate scales. The maximum is reached for relatively low temporal coordinate scales, indicating that strong regularization along the temporal dimension is beneficial for reconstruction quality. For very low temporal coordinate scales, the implicit bias of the network towards slow temporal changes becomes so strong that the SER decreases. 

Figure~\ref{fig:temporal-coordinate-scales-epochs} presents the number of training epochs to reach the maximum SER as a function of the temporal coordinate scale. Similar to the trend observed for spatial coordinate scales, the number of epochs increases as the temporal coordinate scale decreases. 

\begin{figure}[h]
    \centering
    \begin{subfigure}[t]{0.57\textwidth}
        \input{figures/temporal-coordinate-scale-avg-intensity}
        \caption{The graphs are obtained by averaging the pixel intensities within the image patch shown in Figure~\ref{fig:temporal-coordinate-scale-image-patch}. The plot shows that the value of the temporal coordinate scale $\cst$ affects the smoothness of pixel intensities over time. Hence, the regularization strength along the temporal dimension can be adjusted by the temporal coordinate scale.}
        \label{fig:temporal-coordinate-scale-avg-intensity}
    \end{subfigure}
    \hfill
    \begin{subfigure}[t]{0.37\textwidth}
        \centering
\begin{tikzpicture}

\definecolor{darkgray176}{RGB}{176,176,176}

\begin{axis}[
scaled x ticks=manual:{}{\pgfmathparse{#1}},
scaled y ticks=manual:{}{\pgfmathparse{#1}},
width=6cm,
height=6cm,
tick align=outside,
x grid style={darkgray176},
xmajorticks=false,
xmin=0, xmax=134,
xtick style={color=black},
xticklabels={},
y dir=reverse,
y grid style={darkgray176},
ymajorticks=false,
ymin=0, ymax=134,
ytick style={color=black},
yticklabels={},
]
\addplot graphics [includegraphics cmd=\pgfimage, xmin=0, xmax=134, ymin=134, ymax=0] {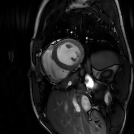};
\draw [red,thick] (100,114) rectangle (120,94);

\end{axis}

\end{tikzpicture}
        \vspace{0.5em}
        \caption{For Figure~\ref{fig:temporal-coordinate-scale-avg-intensity}, we average the pixel intensity of the image patch marked in red. The patch is chosen from a region that is not notably affected by the cardiac and respiratory motion.}
        \label{fig:temporal-coordinate-scale-image-patch}
    \end{subfigure}
    \vspace{0.5em}
    \caption{The temporal coordinate scale $\cst$ affects the temporal smoothness and acts as a regularization parameter for temporal changes.}
\end{figure}

Figure~\ref{fig:temporal-coordinate-scale-avg-intensity} shows how the temporal coordinate scale controls the smoothness of the reconstruction in the temporal dimensions. It plots the average pixel intensity over time within a small image patch that does not contain moving body parts. For low temporal coordinate scales, the average pixel intensity varies smoothly over time, whereas for large scales, e.g., $\cst = \SI{20}{\second^{-1}}$, the average pixel intensity is fluctuating rapidly. The fluctuations can also be seen in the reconstructed video as flickering. The flickering may be caused by the sampling pattern of the low-resolution high-SNR dataset. The image contrast is predominantly contained around the k-space origin that is sampled only in some frames. Thus, the contrast information in the measurement data varies over the frames. 

\FloatBarrier
\subsection{Network size: number of layers and layer width}

The MLP's network size is a critical parameter that affects the reconstruction quality and training time. In this section, we perform an ablation study on the number of hidden layers $\hiddenlayers$ and the number of neurons per hidden layer. 
The reported training time is measured by the wall clock time and thus includes data-loading, logging, and computation of the SER after every epoch.

\begin{figure}[h]
    \centering
    \begin{subfigure}[t]{0.47\textwidth}
\begin{tikzpicture}

\definecolor{crimson2143940}{RGB}{214,39,40}
\definecolor{darkgray176}{RGB}{176,176,176}
\definecolor{darkorange25512714}{RGB}{255,127,14}
\definecolor{forestgreen4416044}{RGB}{44,160,44}
\definecolor{lightgray204}{RGB}{204,204,204}
\definecolor{mediumpurple148103189}{RGB}{148,103,189}
\definecolor{steelblue31119180}{RGB}{31,119,180}

\begin{axis}[
legend cell align={left},
legend style={
  fill opacity=0.8,
  draw opacity=1,
  text opacity=1,
  at={(0.97,0.97)},
  anchor=north east,
  draw=lightgray204
},
tick align=outside,
tick pos=left,
width=\figwidthhalf-3pt,
height=5cm,
x grid style={darkgray176},
xlabel={number of hidden layers $\hiddenlayers$},
xmajorgrids,
xmin=3, xmax=9,
xtick style={color=black},
xtick={3, 5, 7, 9},
y grid style={darkgray176},
ylabel={SER},
ymajorgrids,
ymin=16.9, ymax=17.2,
ytick style={color=black}
]
\addplot [semithick, steelblue31119180, mark=*, mark size=3, mark options={solid}]
table {%
    3 16.96
    5 17.04
    7 17.05
    9 17.08
};


\end{axis}

\end{tikzpicture}
        \caption{The FMLP's image quality in terms of the SER is displayed for different numbers of hidden layers $\hiddenlayers$. Within the examined range, the SER generally improves with more hidden layers.}
        \label{fig:num-layers-SER}
    \end{subfigure}
    \hfill
    \begin{subfigure}[t]{0.47\textwidth}
\begin{tikzpicture}

\definecolor{crimson2143940}{RGB}{214,39,40}
\definecolor{darkgray176}{RGB}{176,176,176}
\definecolor{darkorange25512714}{RGB}{255,127,14}
\definecolor{forestgreen4416044}{RGB}{44,160,44}
\definecolor{lightgray204}{RGB}{204,204,204}
\definecolor{mediumpurple148103189}{RGB}{148,103,189}
\definecolor{steelblue31119180}{RGB}{31,119,180}

\begin{axis}[
legend cell align={left},
legend style={
  fill opacity=0.8,
  draw opacity=1,
  text opacity=1,
  at={(0.01,0.99)},
  anchor=north west,
  draw=lightgray204
},
tick align=outside,
tick pos=left,
width=\figwidthhalf-12pt,
height=5cm,
x grid style={darkgray176},
xlabel={epochs},
xmajorgrids,
xmin=0, xmax=1000,
xtick style={color=black},
y grid style={darkgray176},
ylabel={training time in \SI{}{\minute}},
ymajorgrids,
ymin=0, ymax=300,
ytick style={color=black}
]
\addplot [semithick, steelblue31119180, mark=*, mark size=3, mark options={solid}]
table {%
    958 191
};
\addlegendentry{$3$}
\addplot [thick, lightgray204, no markers, forget plot]
table {%
    0 0
    1000 199.3
};

\addplot [semithick, darkorange25512714, mark=square*, mark size=3, mark options={solid}]
table {%
    449 130
};
\addlegendentry{$5$}
\addplot [thick, lightgray204, no markers, forget plot]
table {%
    0 0
    1000 290
};

\addplot [semithick, forestgreen4416044, mark=triangle*, mark size=3, mark options={solid}]
table {%
    487 194
};
\addlegendentry{$7$}
\addplot [thick, lightgray204, no markers, forget plot]
table {%
    0 0
    1000 398
};

\addplot [semithick, mediumpurple148103189, mark=diamond*, mark size=3, mark options={solid}]
table {%
    570 283
};
\addlegendentry{$9$}
\addplot [thick, lightgray204, no markers, forget plot]
table {%
    0 0
    1000 496
};

\end{axis}

\end{tikzpicture}
        \caption{The training time as measured by the wall-clock time is plotted vs. the number of epochs for different numbers of hidden layers $\hiddenlayers$. The slope of the gray lines indicates the training time per epoch. The number of epochs until the maximum SER is reached is minimized for $\hiddenlayers=5$. Both the training time and the number of epochs increase for fewer or more hidden layers.}
        \label{fig:num-layers-compute-time}
    \end{subfigure}
    \vspace{0.5em}
    \caption{Reconstruction quality in SER a training time as a function of the number of hidden layers $\hiddenlayers$. The experiments are conducted with 512 neurons per layer.}
    \label{fig:num-layers}
\end{figure}
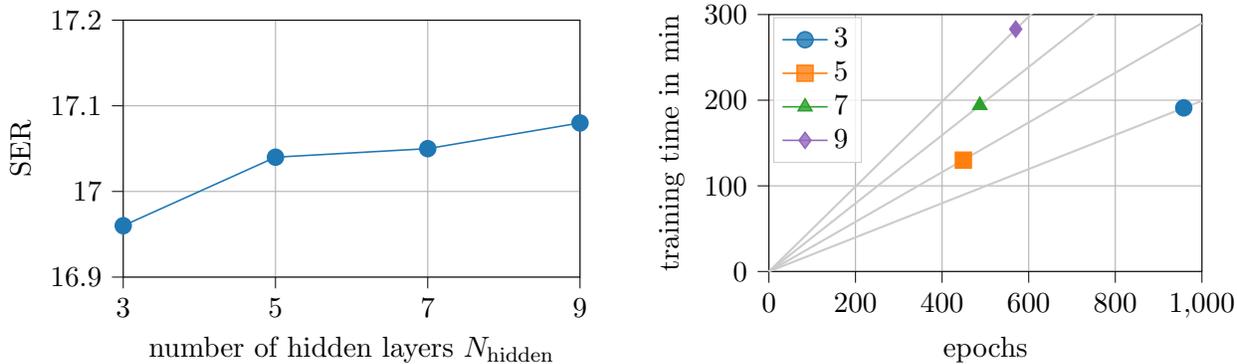

\textbf{Performance as a function of the number of hidden layers: }
We train models with different numbers of hidden layers and measure SER, training time, and number of epochs. 
Figure~\ref{fig:num-layers-SER} shows that on this dataset, the SER increases in increasing the number of layers in the range from 3 to 9 layers.

However, an increased number of hidden layers comes at the cost of a longer training time per epoch as can be seen by the slope of the gray lines in Figure~\ref{fig:num-layers-compute-time} that displays the training time vs. number of epochs for different parameter configurations of $\hiddenlayers$. The shortest training time is achieved for a relatively small number of hidden layers, i.e., $\hiddenlayers = 5$. For more hidden layers, i.e., $\hiddenlayers > 5$, the number of epochs and the training time per epoch increases. For fewer hidden hidden layers, the number of epochs increases substantially, eliminating the benefit of the decreased training time per epoch.

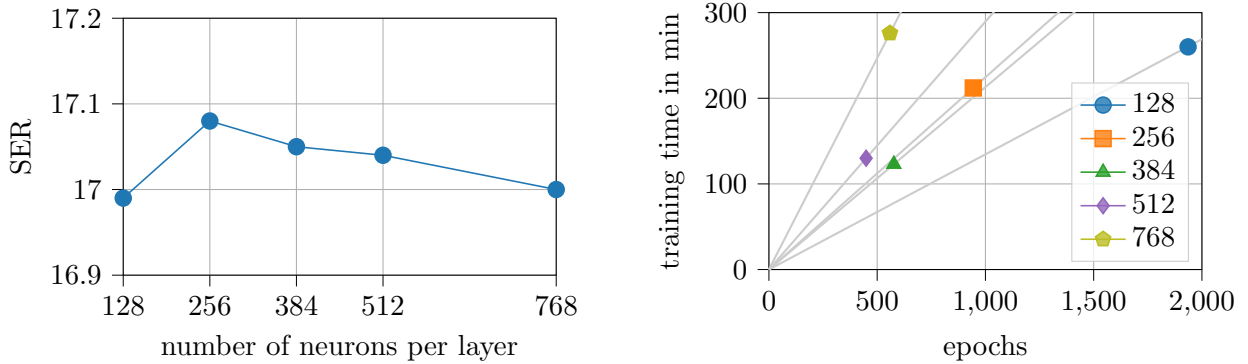
\begin{figure}[h]
    \centering
    \begin{subfigure}[t]{0.47\textwidth}
\begin{tikzpicture}

\definecolor{crimson2143940}{RGB}{214,39,40}
\definecolor{darkgray176}{RGB}{176,176,176}
\definecolor{darkorange25512714}{RGB}{255,127,14}
\definecolor{forestgreen4416044}{RGB}{44,160,44}
\definecolor{lightgray204}{RGB}{204,204,204}
\definecolor{mediumpurple148103189}{RGB}{148,103,189}
\definecolor{steelblue31119180}{RGB}{31,119,180}

\begin{axis}[
legend cell align={left},
legend style={
  fill opacity=0.8,
  draw opacity=1,
  text opacity=1,
  at={(0.97,0.97)},
  anchor=north east,
  draw=lightgray204
},
tick align=outside,
tick pos=left,
width=\figwidthhalf-12pt,
height=5cm,
x grid style={darkgray176},
xlabel={number of neurons per layer},
xmajorgrids,
xmin=128, xmax=768,
xtick style={color=black},
xtick={128, 256, 384, 512, 768},
y grid style={darkgray176},
ylabel={SER},
ymajorgrids,
ymin=16.9, ymax=17.2,
ytick style={color=black}
]
\addplot [semithick, steelblue31119180, mark=*, mark size=3, mark options={solid}]
table {%
    128 16.99
    256 17.08
    384 17.05
    512 17.04
    768 17.00
};

\end{axis}

\end{tikzpicture}
        \caption{The image quality in terms of the SER is plotted for different numbers of neurons per hidden layer. The SER is maximized for 256 neurons per hidden layer. The performance sharply declines for fewer neurons per layer and degrades slowly for an increased number of neurons per layer.}
        \label{fig:layer-width-SER}
    \end{subfigure}
    \hfill
    \begin{subfigure}[t]{0.47\textwidth}
\begin{tikzpicture}

\definecolor{crimson2143940}{RGB}{214,39,40}
\definecolor{darkgray176}{RGB}{176,176,176}
\definecolor{darkorange25512714}{RGB}{255,127,14}
\definecolor{forestgreen4416044}{RGB}{44,160,44}
\definecolor{lightgray204}{RGB}{204,204,204}
\definecolor{mediumpurple148103189}{RGB}{148,103,189}
\definecolor{lime}{RGB}{188,189,34}

\definecolor{steelblue31119180}{RGB}{31,119,180}

\begin{axis}[
legend cell align={left},
legend style={
  fill opacity=0.8,
  draw opacity=1,
  text opacity=1,
  at={(0.97,0.03)},
  anchor=south east,
  draw=lightgray204
},
tick align=outside,
tick pos=left,
width=\figwidthhalf-12pt,
height=5cm,
x grid style={darkgray176},
xlabel={epochs},
xmajorgrids,
xmin=0, xmax=2000,
xtick style={color=black},
y grid style={darkgray176},
ylabel={training time in \SI{}{\minute}},
ymajorgrids,
ymin=0, ymax=300,
ytick style={color=black}
]
\addplot [semithick, steelblue31119180, mark=*, mark size=3, mark options={solid}]
table {%
    1936 260
};
\addlegendentry{128}
\addplot [thick, lightgray204, no markers, forget plot]
table {%
    0 0
    2000 269
};

\addplot [semithick, darkorange25512714, mark=square*, mark size=3, mark options={solid}]
table {%
    944 212
};
\addlegendentry{256}
\addplot [thick, lightgray204, no markers, forget plot]
table {%
    0 0
    2000 449
};

\addplot [semithick, forestgreen4416044, mark=triangle*, mark size=3, mark options={solid}]
table {%
    577 123
};
\addlegendentry{384}
\addplot [thick, lightgray204, no markers, forget plot]
table {%
    0 0
    2000 426
};

\addplot [semithick, mediumpurple148103189, mark=diamond*, mark size=3, mark options={solid}]
table {%
    449 130
};
\addlegendentry{512}
\addplot [thick, lightgray204, no markers, forget plot]
table {%
    0 0
    2000 579
};

\addplot [semithick, lime, mark=pentagon*, mark size=3, mark options={solid}]
table {%
    559 276
};
\addlegendentry{768}
\addplot [thick, lightgray204, no markers, forget plot]
table {%
    0 0
    2000 987
};

\end{axis}

\end{tikzpicture}
        \caption{The training time and the number of epochs are shown for different numbers of neurons per hidden layer. As expected, the training time per epoch increases with the number of neurons per hidden layer (slope of the gray lines), except for the outlier at 256 neurons per layer. The maximum SER is reached with the least amount of training time for 384 neurons per layer.}
        \label{fig:layer-width-compute-time}
    \end{subfigure}
    \vspace{0.5em}
    \caption{Effects of changing the number of neurons per hidden layer. For the experiment, the number of hidden layers is set to $\hiddenlayers = 5$.}
    \label{fig:layer-width}
\end{figure}

\textbf{Performance as a function of the number of neurons per hidden layer: }
We now consider the number of neurons per hidden layer and report the SER, training time, and the number of epochs. In Figure~\ref{fig:layer-width-SER}, the SER is plotted as a function of the number of neurons per hidden layer. The maximum SER is achieved for a relatively low number of neurons (256) and slightly decreases for increasing number of neurons. For fewer neurons, i.e., 128, the SER declines rapidly, presumably as the network can no longer fit the target signal.

Figure~\ref{fig:layer-width-compute-time} shows the training time and the number of epochs for different numbers of neurons per layer. As expected, the training time per epoch increases in the number of neurons. The shortest overall training time is achieved for 384 neurons per layer, i.e., for moderate numbers of neurons. For an increased number of neurons, the longer training time per epoch mainly affects the overall training time. For fewer neurons, the increased number of epochs compensates the benefits of the decreased training time per epoch.

\FloatBarrier
\subsection{Explicit temporal TV regularization}
\label{subsec:explicit-tv-regularization}

We study if the reconstruction quality of the FMLP can be improved by adding an additional loss term that explicitly regularizes the temporal relation by imposing a total variation (TV) norm penalty. We find that such regularization does not notably improve the image quality if the temporal coordinate scale is tuned properly.

For the experiment, we add a TV norm penalty to the training loss
\begin{equation*}
     \loss(\fmlpparams) = \frac{1}{\frames} \sum\limits_{\frameindex=1}^{\frames} \norm{\measop_{\frameindex} \fmlpgrid(t_{\frameindex}) - \meas_{\frameindex}}_2^2 + \lambda_{\text{TV}} \loss_{\text{TV}, \frameindex}(\fmlpparams),
\end{equation*}
where $\loss_{\text{TV}, \frameindex} > 0$ is a hyperparameter that adjusts the regularization strength, and the total-variation loss is defined as follows. We compute the temporal TV loss between the current reconstruction $\fmlpgrid(t_{\frameindex})$ and the latest reconstructions $\imgrecold_i, \, i=1,\dots, \frames$ that are stored form an earlier training epoch or from an earlier iteration within the current epoch. The batch-size is set to 1, meaning that gradients are computed for individual frames. The gradient of the loss is only computed with respect to the current reconstruction $\fmlpgrid(t_{\frameindex})$. The loss computes as 
\begin{equation*}
    \loss_{\text{TV}, \frameindex}(\fmlpparams) = 
    \begin{cases}
        \norm{\fmlpgrid(t_1) - \imgrecold_2}_1, & \frameindex = 1,\\
        \norm{\fmlpgrid(t_{\frameindex}) - \imgrecold_{\frameindex-1}}_1 + \norm{\fmlpgrid(t_{\frameindex}) - \imgrecold_{\frameindex+1}}_1, & \frameindex < 1 \wedge  \frameindex < \frames, \\
        \norm{\fmlpgrid(t_{\frames}) - \imgrecold_{\frames-1}}_1, & \frameindex = \frames,
    \end{cases}
\end{equation*}
where the $l_1$ norm is computed for the real and imaginary part separately and is added.

\begin{figure}[h]
    \centering
\begin{tikzpicture}

\definecolor{crimson2143940}{RGB}{214,39,40}
\definecolor{darkgray176}{RGB}{176,176,176}
\definecolor{darkorange25512714}{RGB}{255,127,14}
\definecolor{forestgreen4416044}{RGB}{44,160,44}
\definecolor{lightgray204}{RGB}{204,204,204}
\definecolor{mediumpurple148103189}{RGB}{148,103,189}
\definecolor{steelblue31119180}{RGB}{31,119,180}

\begin{semilogxaxis}[
legend cell align={left},
legend style={
  fill opacity=0.8,
  draw opacity=1,
  text opacity=1,
  at={(0.03,0.03)},
  anchor=south west,
  draw=lightgray204
},
tick align=outside,
tick pos=left,
width=10cm,
height=6cm,
x grid style={darkgray176},
xlabel={$\lambda_{\text{TV}}$},
xmajorgrids,
xmin=1, xmax=1000,
xtick style={color=black},
y grid style={darkgray176},
ylabel={SER},
ymajorgrids,
ymin=16.4, ymax=17.25,
ytick style={color=black}
]
\addplot [semithick, steelblue31119180, mark=*, mark size=3, mark options={solid}]
table {%
    1 17.07
    10 17.11
    100 17.11
    1000 16.51
};
\addlegendentry{$\cst = \SI{5.3}{\second^{-1}}$}

\addplot [semithick, steelblue31119180, dashed]
    table {%
        1 17.05
        1000 17.05
    };
\addlegendentry{$\cst = \SI{5.3}{\second^{-1}}$, $\lambda_{\text{TV}} = 0$}

\addplot [semithick, darkorange25512714, mark=square*, mark size=3, mark options={solid}]
table {%
    0 17.16
    10 17.17
    100 17.17
    1000 16.45
};
\addlegendentry{$\cst = \SI{1.0}{\second^{-1}}$}

\addplot [semithick, darkorange25512714, dashed]
    table {%
        1 17.16
        1000 17.16
    };
\addlegendentry{$\cst = \SI{1.0}{\second^{-1}}$, $\lambda_{\text{TV}} = 0$}

\end{semilogxaxis}

\end{tikzpicture}
    \vspace{0.5em}
    \caption{The image quality as measured by the SER is plotted for different choices of the hyperparameter $\lambda_{\text{TV}}$ that controls the strength of the explicit TV regularization. The explicit temporal TV regularization improves the SER for higher spatial coordinate scales, i.e., weak implicit regularization. If the implicit regularization is stronger, i.e., for lower coordinate scales, the SER does not improve much by regularizing explicitly. For both scenarios, the SER drops rapidly if the explicit temporal regularization is too strong.}
    \label{fig:temporal-tv-regularization}
\end{figure}
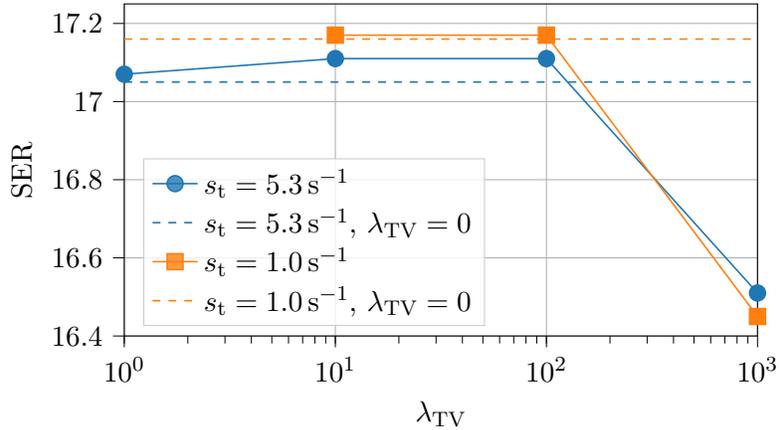

Figure~\ref{fig:temporal-tv-regularization} shows the reconstruction quality as measured by the SER as a function of the hyperparameter $\lambda_{\text{TV}}$ for different choices of the temporal coordinate scale $\cst$. It can be seen that the SER does not improve notably for regulariztion strength beyond $\lambda_{\text{TV}} = 0$ (no regularization), when the implicit regularization is sufficiently well tuned.  Therefore, additional TV regularization is not recommended in the optimized operating regime of the FMLP. 

\subsection{Separate spatial and temporal Fourier-feature embeddings yields sharper images}
\label{subsec:combined-embedding}
Our proposed network architecture uses separate Fourier-feature embeddings for the spatial coordinates and the temporal coordinate. An alternative is to jointly embed the spatial and temporal coordinates. In this section, we compare the performance of FMLP with separate and joint embeddings. 

As before, the separate embedding is as explained in the main document. 
For the joint embedding, the spatio-temporal coordinate vector is embedded in the  640 dimensional feature vector
\begin{equation*}
    \fmap_{\text{joint}}([x,y,t]) 
    = \left[\sin\left(\tilde{\fmapwavematrix}\begin{bmatrix}\csx x\\ \csy y\\ \cst t\end{bmatrix}\right),\, \cos\left(\tilde{\fmapwavematrix}\begin{bmatrix} \csx x \\ \csy y\\ \cst t\end{bmatrix}\right) \right]^T,
\end{equation*}
where the elements in the matrix $\tilde{\fmapwavematrix} \in \realnumbers^{320 \times 3}$ are drawn independently from the standard normal distribution. The coordinate scales are set to the their default values for both configurations.

\begin{figure}[h]
    \centering
\begin{tikzpicture}

\newcommand\imgwidth[0]{5cm}

\definecolor{darkgray176}{RGB}{176,176,176}

\begin{groupplot}[
  group style={
    group size=3 by 1,
    ylabels at=edge left,
    xlabels at=edge top,
    vertical sep=0.1cm,
    horizontal sep=0.1cm
  }]

\nextgroupplot[
scaled x ticks=manual:{}{\pgfmathparse{#1}},
scaled y ticks=manual:{}{\pgfmathparse{#1}},
width=\imgwidth,
height=\imgwidth,
tick align=outside,
x grid style={darkgray176},
xmajorticks=false,
xmin=0, xmax=200,
xtick style={color=black},
xticklabels={},
y dir=reverse,
y grid style={darkgray176},
ymajorticks=false,
ymin=0, ymax=200,
ytick style={color=black},
yticklabels={},
ylabel={},
xlabel={separate embeddings}
]
\addplot graphics [includegraphics cmd=\pgfimage, xmin=0, xmax=200, ymin=200, ymax=0] {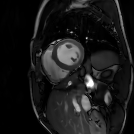};

\nextgroupplot[
scaled x ticks=manual:{}{\pgfmathparse{#1}},
scaled y ticks=manual:{}{\pgfmathparse{#1}},
width=\imgwidth,
height=\imgwidth,
tick align=outside,
x grid style={darkgray176},
xmajorticks=false,
xmin=0, xmax=200,
xtick style={color=black},
xticklabels={},
y dir=reverse,
y grid style={darkgray176},
ymajorticks=false,
ymin=0, ymax=200,
ytick style={color=black},
yticklabels={},
xlabel={joint embedding}
]
\addplot graphics [includegraphics cmd=\pgfimage, xmin=0, xmax=200, ymin=200, ymax=0] {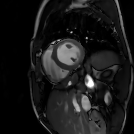};

\nextgroupplot[
scaled x ticks=manual:{}{\pgfmathparse{#1}},
scaled y ticks=manual:{}{\pgfmathparse{#1}},
width=\imgwidth,
height=\imgwidth,
tick align=outside,
x grid style={darkgray176},
xmajorticks=false,
xmin=0, xmax=200,
xtick style={color=black},
xticklabels={},
y dir=reverse,
y grid style={darkgray176},
ymajorticks=false,
ymin=0, ymax=200,
ytick style={color=black},
yticklabels={},
xlabel={BH reference}
]
\addplot graphics [includegraphics cmd=\pgfimage, xmin=0, xmax=200, ymin=200, ymax=0] {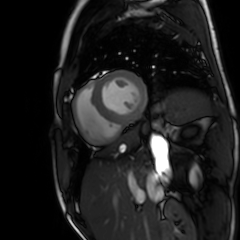};

\end{groupplot}

\end{tikzpicture}
    \vspace{0.5em}
    \caption{Reconstructions by the FMLP are shown for architectures with separate and joint embedding of the spatial and temporal coordinates, alongside with a reconstruction of the low-resolution high-SNR dataset breath-hold dataset that serves as visual reference. The architecture with separate coordinate embeddings yields an improved image quality with sharper details compared to the architecture with a joint embedding.The reconstructions are shown for frame $\frameindex = 106$.}
    \label{fig:combined-embedding-imgs}
\end{figure}

Our results show that the image quality in terms of the SER drops from $17.05$ to $16.76$ if the coordinates are embedded jointly. The difference is also visible in Figure~\ref{fig:combined-embedding-imgs}, where separating embeddings of the spatial and temporal coordinates yield sharper reconstructions than joint embeddings.

\FloatBarrier
\subsection{Temporal resolution of the binning into frames}
Our measurement model is based on the assumption that the heart is essentially static during a frame. However, even though the frame duration is very small, the heart is moving continuously while measurements are taken sequentially.
The longer the frame, the more the heart moves during the frame, and the larger the error in assuming the heart is constant. 
In this section, we perform an ablation study over the duration of frames, i.e., the temporal resolution of the binning and study if the image quality improves when improving the temporal resolution.

The temporal resolution is determined by the number of measured $\ky$-lines $\numlines$ that are binned together in a frame. We train the FMLP on datasets with different $\numlines$, ranging from $4$ to $36$ and report the image quality as measured by the SER and the training time. Note that we change the temporal resolution of the frames without significantly changing the amount of training data. As we increase the number of lines per frame, the number of frames is decreased, such that the total number of measurements stays approximately constant at $225 \cdot 6 = 1350$ lines. We extract the same lines for validation across all temporal resolutions such that the SER values can be compared across temporal resolutions.

\begin{figure}[h]
    \centering
    \begin{subfigure}[t]{0.47\textwidth}
\begin{tikzpicture}

\definecolor{crimson2143940}{RGB}{214,39,40}
\definecolor{darkgray176}{RGB}{176,176,176}
\definecolor{darkorange25512714}{RGB}{255,127,14}
\definecolor{forestgreen4416044}{RGB}{44,160,44}
\definecolor{lightgray204}{RGB}{204,204,204}
\definecolor{mediumpurple148103189}{RGB}{148,103,189}
\definecolor{steelblue31119180}{RGB}{31,119,180}

\begin{axis}[
legend cell align={left},
legend style={
  fill opacity=0.8,
  draw opacity=1,
  text opacity=1,
  at={(0.97,0.97)},
  anchor=north east,
  draw=lightgray204
},
tick align=outside,
tick pos=right,
width=\figwidthhalf-12pt,
height=5cm,
x grid style={darkgray176},
xlabel={temporal resolution in \SI{}{\milli \second}},
xmajorgrids,
xmin=4, xmax=36,
xtick style={color=black},
xtick={4, 6, 8, 12, 18, 24, 36},
xticklabels={11.2, 16.8, 22.4, 33.6, 50.4, 67.2, 100.8},
y grid style={darkgray176},
ylabel={},
xticklabel style={rotate=60},
ymajorticks=false,
ymin=16.8, ymax=17.1,
ytick style={color=black}
]
\end{axis}

\begin{axis}[
legend cell align={left},
legend style={
  fill opacity=0.8,
  draw opacity=1,
  text opacity=1,
  at={(0.97,0.97)},
  anchor=north east,
  draw=lightgray204
},
tick align=outside,
tick pos=left,
width=\figwidthhalf-12pt,
height=5cm,
x grid style={darkgray176},
xlabel={number of lines per frame $\numlines$},
xmajorgrids,
xmin=4, xmax=36,
xtick style={color=black},
xtick={4, 6, 8, 12, 18, 24, 36},
y grid style={darkgray176},
ylabel={SER},
ymajorgrids,
ymin=16.8, ymax=17.1,
ytick style={color=black}
]
\addplot [semithick, steelblue31119180, mark=*, mark size=3, mark options={solid}]
table {%
    4 17.07
    6 17.05
    8 17.04
    12 17.04
    18 16.99
    24 16.96
    36 16.87
};

\end{axis}

\end{tikzpicture}
        \caption{The FMLP is trained on datasets with binnings of different temporal resolutions, i.e., different number of lines per frame $\numlines$, and the SER is computed on a validation dataset that is kept constant for all temporal resolutions. The figure shows that the image quality improves with increased temporal resolution, i.e., lower $\numlines$.}
        \label{fig:num-lines-per-frame-SER}
    \end{subfigure}
    \hfill
    \begin{subfigure}[t]{0.47\textwidth}
\begin{tikzpicture}

\definecolor{crimson2143940}{RGB}{214,39,40}
\definecolor{darkgray176}{RGB}{176,176,176}
\definecolor{darkorange25512714}{RGB}{255,127,14}
\definecolor{forestgreen4416044}{RGB}{44,160,44}
\definecolor{lightgray204}{RGB}{204,204,204}
\definecolor{mediumpurple148103189}{RGB}{148,103,189}
\definecolor{steelblue31119180}{RGB}{31,119,180}

\begin{axis}[
legend cell align={left},
legend style={
  fill opacity=0.8,
  draw opacity=1,
  text opacity=1,
  at={(0.97,0.97)},
  anchor=north east,
  draw=lightgray204
},
tick align=outside,
tick pos=left,
width=\figwidthhalf-3pt,
height=5cm,
x grid style={darkgray176},
xlabel={number of lines per frame $\numlines$},
xmajorgrids,
xmin=4, xmax=36,
xtick style={color=black},
xtick={4, 6, 8, 12, 18, 24, 36},
y grid style={darkgray176},
ylabel={training time in \SI{}{\minute}},
ymajorgrids,
ymin=50, ymax=400,
ytick style={color=black}
]
\addplot [semithick, steelblue31119180, mark=*, mark size=3, mark options={solid}]
table {%
    4 349
    6 196
    8 170
    12 111
    18 87
    24 96
    36 63
};

\end{axis}

\end{tikzpicture}
        \caption{The training time measured by the wall-clock time is reported for different temporal resolutions, i.e., different numbers of $\ky$-lines per frame. The training time increases with decreasing $\numlines$ due to the increasing number of frames $\frames$ that need to be processed in each epoch.}
        \label{fig:num-lines-per-frame-compute-time}
    \end{subfigure}
    \vspace{0.5em}
    \caption{Increasing the temporal resolution of the frames improves the maximum SER but comes at the cost of increased training time.}
\end{figure}
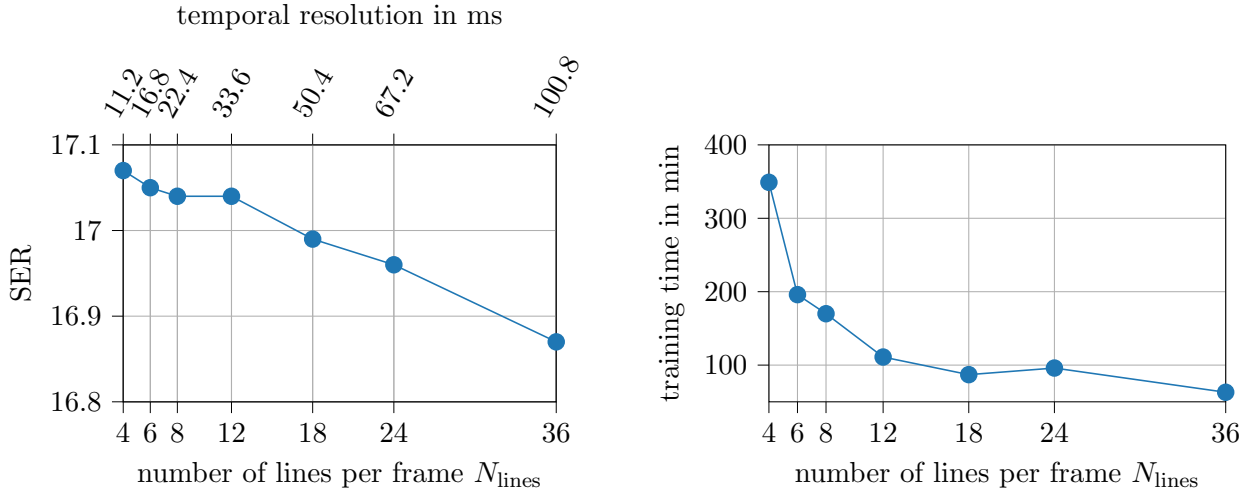

Figure~\ref{fig:num-lines-per-frame-SER} shows that the SER improves as the number of measured k-space lines per frame $\numlines$ decreases, which is equivalent to increasing the temporal resolution. 
The result is to be expected as the mismatch between the true measurement process and our measurement model is reduced by decreasing $\numlines$. With increased temporal resolution, the motion within a frame is reduced.
Increasing the temporal resolution, however, comes at the cost of longer training time as the model has the be trained on more frames per epoch, see Figure~\ref{fig:num-lines-per-frame-compute-time}.

\FloatBarrier
\section{Ablation Studies of the KFMLP}

We perform ablation studies on several of the KFMLP's hyperparameters including the spatial coordinate scales $\csx, \csy$, the output scale $\sout$ of the network, the hyperparameter $\epsilon$ of the high-dynamic-range loss, and the hyperparameters $\lambdadenoiser$ and $\sigma$ of the k-space denoiser.
The ablation studies are conducted on the low-resolution high-SNR dataset.
The KFMLP is trained until no new SER high-score has been achieved in the last 200 epochs. A total of $\frames=225$ frames of data are used for the reconstruction with 6 $\ky$-lines per frame. It is important to note that the bin size does not directly affect the temporal resolution of the reconstructions since the KFMLP can be fitted at the exact temporal coordinates without binning. However, the number of $\ky$-lines per frame is the equivalent of a batch-size, and therefore affects the computational speed. 

For brevity, we define a default hyperparameter configuration. Unless not stated explicitly, the hyper-parameters are: $\sout = 1000$, $\cst = \SI{1.0}{\second^{-1}}$, $\csx = \csy = 10$, 7 hidden layers with 512 neurons per layer.

\subsection{SER is sensitive to the choice of the spatial coordinate scales}
\label{subsec:kfmlp_sx}
\begin{figure}[h]
    \centering
\begin{tikzpicture}

\definecolor{crimson2143940}{RGB}{214,39,40}
\definecolor{darkgray176}{RGB}{176,176,176}
\definecolor{darkorange25512714}{RGB}{255,127,14}
\definecolor{forestgreen4416044}{RGB}{44,160,44}
\definecolor{lightgray204}{RGB}{204,204,204}
\definecolor{mediumpurple148103189}{RGB}{148,103,189}
\definecolor{steelblue31119180}{RGB}{31,119,180}

\begin{axis}[
legend cell align={left},
legend style={
  fill opacity=0.8,
  draw opacity=1,
  text opacity=1,
  at={(0.97,0.97)},
  anchor=north east,
  draw=lightgray204
},
log basis x={10},
tick align=outside,
tick pos=left,
x grid style={darkgray176},
xlabel={$s_{\text{x}} = s_{\text{y}}$},
xmin=3, xmax=100000,
xmode=log,
xtick style={color=black},
y grid style={darkgray176},
ylabel={SER},
ymin=8, ymax=13,
xmajorgrids,
ymajorgrids,
ytick style={color=black}
]

\addplot [semithick, mediumpurple148103189, mark=*]
table {%
3 12.04
10 12.44
100 11.02
};
\addlegendentry{$s_{\text{out}} = 100$}

\addplot [semithick, steelblue31119180, mark=square*]
table {%
3 12
10 12.43
30 10.9
100 8.35
1000 8.35
};
\addlegendentry{$s_{\text{out}} = 1000$}

\addplot [semithick, darkorange25512714, mark=triangle*]
table {%
3 11.87
10 12.4
30 10.94
};
\addlegendentry{$s_{\text{out}} = 3000$}

\addplot [semithick, forestgreen4416044, mark=diamond*]
table {%
100 8.43
1000 8.36
10000 8.32
100000 8.34
};
\addlegendentry{$s_{\text{out}} = 3600$}

\end{axis}

\end{tikzpicture}
    \vspace{0.5em}
    \caption{An ablation study of the KFMLP's spatial coordinate scales and the output scale $\sout$ is conducted. The standard $l_2$ reconstruction loss is used and the k-space denoiser is deactivated by setting $\lambdadenoiser = 0$. The image quality is reported in terms of the SER. The highest SER scores are achieved for relatively low spatial coordinate scales with a sharp maximum at around $\csx = \csy = 10$.}
    \label{fig:kfmlp-sx}
\end{figure}
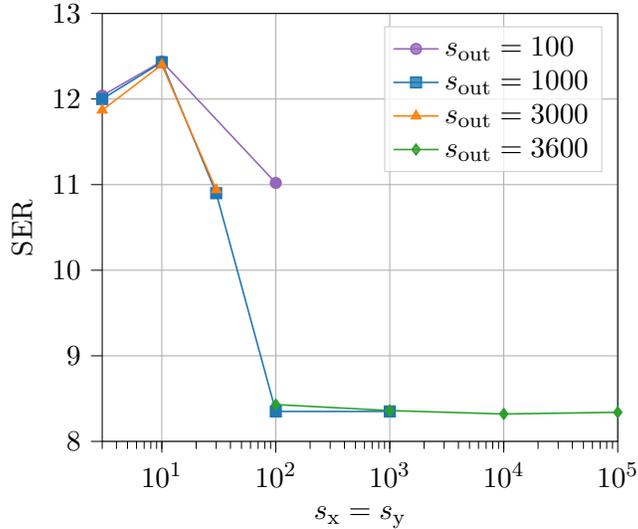

The spatial coordinate scales $\csx$ and $\csy$ affect the implicit bias of the network towards low-frequency signals and need to be tuned carefully. In this section, we perform a grid search over the spatial coordinate scales $\csx = \csy$ and measure the achieved reconstruction quality in SER. The experiment is repeated for different output scales $\sout$. The experiment is conducted with the following $l_2$ reconstruction loss instead of the high-dynamic-range loss, and without the k-space denoiser.
\begin{equation*}
    \loss(\fmlpparams) = \frac{1}{\frames}\sum\limits_{\frameindex=1}^{\frames}\norm{\kfmlptrajectory[,\frameindex] -  \meas_{\frameindex}}_2^2
\end{equation*}
By first evaluating the performance using the $l_2$ loss we establish a baseline for comparing the performance to the high-dynamic-range loss. 
Figure~\ref{fig:kfmlp-sx} illustrates the SER values obtained for different spatial coordinate scales. The figure demonstrates that the KFMLP's reconstruction performance is significantly affected by variations in spatial coordinate scales. Among the tested configurations, the maximum SER is achieved for $\csx = \csy = 10$ and declines rapidly for lower and higher coordinate scales. Thus, careful tuning is necessary to ensure good performance.

\subsection{High-dynamic-range reconstruction loss improves the SER marginally}
\label{subsec:hdr-loss}
In this section, the KFMLP is trained using the HDR reconstruction loss and an ablation study on the hyperparameter $\epsilon$ of the HDR loss is performed. The k-space denoiser is disabled by setting $\lambdadenoiser = 0$. Thereby, the performance of the KFMLP with HDR loss can be compared to the performance with the $l_2$ reconstruction loss that was used in the previous Section~\ref{subsec:kfmlp_sx}.
The study is repeated for two different spatial coordinate scales $\csx=\csy=3$ and $\csx=\csy=10$.

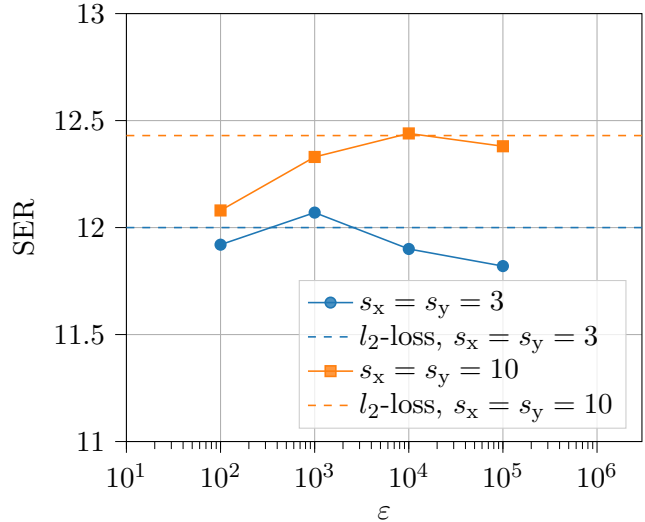
\begin{figure}[h]
    \centering
\begin{tikzpicture}

\definecolor{crimson2143940}{RGB}{214,39,40}
\definecolor{darkgray176}{RGB}{176,176,176}
\definecolor{darkorange25512714}{RGB}{255,127,14}
\definecolor{forestgreen4416044}{RGB}{44,160,44}
\definecolor{lightgray204}{RGB}{204,204,204}
\definecolor{mediumpurple148103189}{RGB}{148,103,189}
\definecolor{steelblue31119180}{RGB}{31,119,180}

\begin{axis}[
legend cell align={left},
legend style={
  fill opacity=0.8,
  draw opacity=1,
  text opacity=1,
  at={(0.97,0.03)},
  anchor=south east,
  draw=lightgray204
},
log basis x={10},
tick align=outside,
tick pos=left,
x grid style={darkgray176},
xlabel={$\varepsilon$},
xmin=10, xmax=3000000,
xmode=log,
xtick style={color=black},
y grid style={darkgray176},
ylabel={SER},
ymin=11, ymax=13,
xmajorgrids,
ymajorgrids,
ytick style={color=black}
]
\addplot [semithick, steelblue31119180, mark=*]
table {%
100 11.92
1000 12.07
10000 11.9
100000 11.82
};
\addlegendentry{$s_{\text{x}} = s_{\text{y}} = 3$}

\addplot [semithick, steelblue31119180, dashed]
table {%
10 12
3000000 12
};
\addlegendentry{$l_2$-loss, $s_{\text{x}} = s_{\text{y}} = 3$}

\addplot [semithick, darkorange25512714, mark=square*]
table {%
100 12.08
1000 12.33
10000 12.44
100000 12.38
};
\addlegendentry{$s_{\text{x}} = s_{\text{y}} = 10$}

\addplot [semithick, darkorange25512714, dashed]
table {%
10 12.43
3000000 12.43
};
\addlegendentry{$l_2$-loss, $s_{\text{x}} = s_{\text{y}} = 10$}

\end{axis}

\end{tikzpicture}
    \vspace{0.5em}
    \caption{
    The KFMLP's image quality in terms of the SER is shown for a different choices of the hyperparameter $\varepsilon$ that adjusts the compression of the dynamic range. For comparison, the results of the KFMLP with $l_2$ reconstruction loss are included. 
    The figure shows that the HDR loss slightly improves the SER compared to the $l_2$ loss if $\varepsilon$ is tuned properly. If not tuned properly, the HDR loss performs worse than the standard $l_2$ loss.}
    \label{fig:kfmlp_hdr}
\end{figure}

Figure~\ref{fig:kfmlp_hdr} shows the image quality measured by the SER as a function of the hyperparameter $\varepsilon$. For comparison, the SER with the $l_2$ reconstruction loss is included in the graph. The HDR loss improves the SER only marginally compared to the standard $l_2$ reconstruction loss. If $\varepsilon$ is not tuned properly, however, the HDR loss degrades the reconstruction quality.
Thus, using the HDR loss is only advisable if $\varepsilon$ can be tuned carefully.

\subsection{K-space denoiser improves the SER marginally}
\label{subsec:kspace-denoiser}
In the last section, the KFMLP's hyperparameters were optimized with a disabled k-space denoiser. In this section, the k-space denoiser is enabled and an ablation study of the the denoiser's hyperparameters $\lambdadenoiser$ and $\sigma$ is conducted.

\begin{figure}[h]
    \centering
\begin{tikzpicture}

\definecolor{crimson2143940}{RGB}{214,39,40}
\definecolor{darkgray176}{RGB}{176,176,176}
\definecolor{darkorange25512714}{RGB}{255,127,14}
\definecolor{forestgreen4416044}{RGB}{44,160,44}
\definecolor{lightgray204}{RGB}{204,204,204}
\definecolor{mediumpurple148103189}{RGB}{148,103,189}
\definecolor{steelblue31119180}{RGB}{31,119,180}

\begin{axis}[
legend cell align={left},
legend style={
  fill opacity=0.8,
  draw opacity=1,
  text opacity=1,
  at={(0.97,0.03)},
  anchor=south east,
  draw=lightgray204
},
log basis x={10},
tick align=outside,
tick pos=left,
x grid style={darkgray176},
xlabel={$\sigma$},
xmin=0.01, xmax=1000,
xmode=log,
xtick style={color=black},
y grid style={darkgray176},
ylabel={SER},
ymin=11, ymax=13,
xmajorgrids,
ymajorgrids,
ytick style={color=black}
]

\addplot [semithick, steelblue31119180, mark=*]
table {%
1 12.43
};
\addlegendentry{$\lambda_{\text{denoiser}} = 0.01$}

\addplot [semithick, forestgreen4416044, mark=square*]
table {%
0.01 11.59
0.1 11.73
1 12.47
10 12.51
100 12.42
1000 12.42
};
\addlegendentry{$\lambda_{\text{denoiser}} = 0.1$}

\addplot [semithick, darkorange25512714, mark=triangle*]
table {%
1 12.07
};
\addlegendentry{$\lambda_{\text{denoiser}} = 1$}

\addplot [semithick, darkgray176]
table {%
0.01 12.44
1000 12.44
};
\addlegendentry{no denoiser}

\end{axis}

\end{tikzpicture}
    \vspace{0.5em}
    \caption{The KFMLP's image quality measured by the SER is plotted for different configurations of the k-space denoiser's hyperparameters $\sigma$ and $\lambdadenoiser$. For comparison the performance with disabled k-space denoiser ($\lambdadenoiser = 0$) is included in the graph. The figure shows that the k-space denoiser slightly improves the maximum SER of the KFMLP but not by a substantial margin.}
    \label{fig:kfmlp_denoiser}
\end{figure}
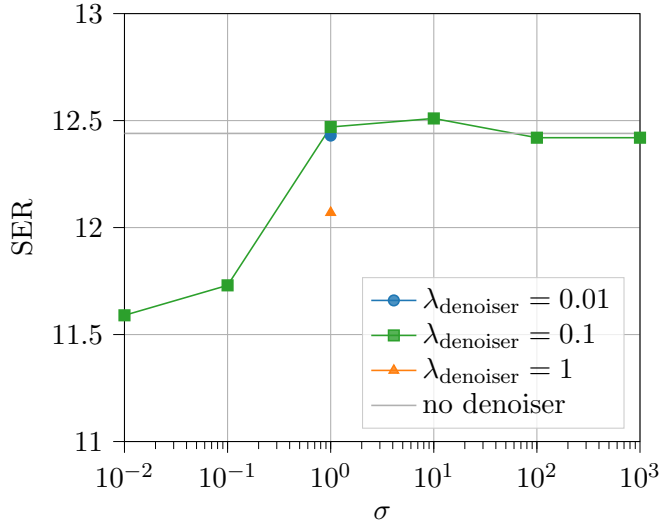

Figure~\ref{fig:kfmlp_denoiser} shows the SER as a function of $\sigma$ controlling the decay of the weights in the filter matrix $\boldsymbol{K}$. The experiment is conducted for different regularization strengths $\lambdadenoiser$.
The figure demonstrates that the k-space denoiser improves the SER if the hyper-parameters $\lambdadenoiser$ and $\sigma$ are tuned properly. However, the improvements are small and are only achieved for a precise tuning of the hyperparameters.

\FloatBarrier
\section{Experiments on Phantom Data}
\label{sec:phantom-experiments}
It is common practice to measure the image quality with full-reference image quality metrics. However, the required reference images are not available for the datasets used in our other experiments. Thus, we conduct additional experiment on phantom data, where ground-truth images are available.

\subsection{Phantom dataset}
\label{subsec:phantom-datasets}
We generated a phantom dataset with MRXCAT \cite{wissmannMRXCATRealisticNumerical2014}, a numerical phantom that can simulate cardiac and respiratory motion. It computes synthetic MR images from body slices of the 4D cardiac-torso phantom XCAT \cite{segars4DXCATPhantom2010}. The synthetic images are passed through the imaging equation to obtain synthetic k-space data. 

We generate datasets with the exact same sampling masks $\mask_{\frameindex}$ as for the experimental low-resolution datasets. Each measured k-space line is synthesized from an individual image in a different motion state to imitate the sequential sampling of k-space coordinates. 

\subsection{Comparison of the methods by evaluating full-reference image quality metrics}
\label{subsec:image-quality-phantom-data}
The FMLP, KFMLP, and the t-DIP are evaluated on the phantom dataset and the full reference image quality metrics SSIM and VIF are computed using the ground-truth images that are available for the phantom dataset. The results are reported in Table~\ref{tab:metrics-phantom}. 
We find that the FMLP achieves marginally higher SSIM and VIF scores on the phantom dataset than the t-DIP.
The SER, by contrast, is slightly higher for the t-DIP than for the FMLP. The differences are small and the two methods give similar reconstruction quality. The KFMLP achieves much lower SER, SSIM, and VIF scores than FMLP and t-DIP.

\begin{table}[h]
    \centering
    \begin{tabular}{c | c | c | c}
    \specialrule{.15em}{.05em}{.05em}
       metric & FMLP & KFMLP & t-DIP \\
        \midrule
        max. SER & 23.70 & 12.73 & \textbf{23.74} \\
        SSIM at max. SER epoch & \textbf{0.986} & 0.126 & 0.973 \\
        VIF at max. SER epoch& \textbf{0.694} & 0.120 & 0.682 \\
    \specialrule{.15em}{.05em}{.05em}
    \end{tabular}
    \vspace{0.5em}
    \caption{The performance metrics are evaluated on the phantom dataset. The t-DIP achieves a higher data consistency according to the SER, however, the FMLP slightly outperforms the t-DIP in the image quality metrics SSIM and VIF. The KFMLP achieves much lower SER, SSIM, and VIF scores. The models were trained on $\frames=225$ frames.}
    \label{tab:metrics-phantom}
\end{table}

\FloatBarrier
\section{Reconstructed Images at Larger Field-of-View}
\label{sec:full-res-images}
\begin{figure}[H]
    \centering
\begin{tikzpicture}

\newcommand\imgwidth[0]{5cm}

\definecolor{darkgray176}{RGB}{176,176,176}

\begin{groupplot}[
  group style={
    group size=4 by 4,
    ylabels at=edge left,
    xlabels at=edge top,
    vertical sep=0.1cm,
    horizontal sep=0.1cm
  }]

\nextgroupplot[
scaled x ticks=manual:{}{\pgfmathparse{#1}},
scaled y ticks=manual:{}{\pgfmathparse{#1}},
width=\imgwidth,
height=\imgwidth,
tick align=outside,
x grid style={darkgray176},
xmajorticks=false,
xmin=0, xmax=200,
xtick style={color=black},
xticklabels={},
y dir=reverse,
y grid style={darkgray176},
ymajorticks=false,
ymin=0, ymax=200,
ytick style={color=black},
yticklabels={},
ylabel={$\frameindex = 106$},
xlabel={FMLP}
]
\addplot graphics [includegraphics cmd=\pgfimage, xmin=0, xmax=200, ymin=200, ymax=0] {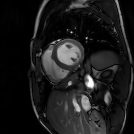};

\nextgroupplot[
scaled x ticks=manual:{}{\pgfmathparse{#1}},
scaled y ticks=manual:{}{\pgfmathparse{#1}},
width=\imgwidth,
height=\imgwidth,
tick align=outside,
x grid style={darkgray176},
xmajorticks=false,
xmin=0, xmax=200,
xtick style={color=black},
xticklabels={},
y dir=reverse,
y grid style={darkgray176},
ymajorticks=false,
ymin=0, ymax=200,
ytick style={color=black},
yticklabels={},
xlabel={KFMLP}
]
\addplot graphics [includegraphics cmd=\pgfimage, xmin=0, xmax=200, ymin=200, ymax=0] {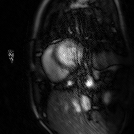};

\nextgroupplot[
scaled x ticks=manual:{}{\pgfmathparse{#1}},
scaled y ticks=manual:{}{\pgfmathparse{#1}},
width=\imgwidth,
height=\imgwidth,
tick align=outside,
x grid style={darkgray176},
xmajorticks=false,
xmin=0, xmax=200,
xtick style={color=black},
xticklabels={},
y dir=reverse,
y grid style={darkgray176},
ymajorticks=false,
ymin=0, ymax=200,
ytick style={color=black},
yticklabels={},
xlabel={t-DIP}
]
\addplot graphics [includegraphics cmd=\pgfimage,xmin=0, xmax=200, ymin=200, ymax=0] {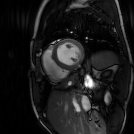};

\nextgroupplot[
scaled x ticks=manual:{}{\pgfmathparse{#1}},
scaled y ticks=manual:{}{\pgfmathparse{#1}},
width=\imgwidth,
height=\imgwidth,
tick align=outside,
x grid style={darkgray176},
xmajorticks=false,
xmin=0, xmax=200,
xtick style={color=black},
xticklabels={},
y dir=reverse,
y grid style={darkgray176},
ymajorticks=false,
ymin=0, ymax=200,
ytick style={color=black},
yticklabels={},
xlabel={BH reference}
]
\addplot graphics [includegraphics cmd=\pgfimage,xmin=0, xmax=200, ymin=200, ymax=0] {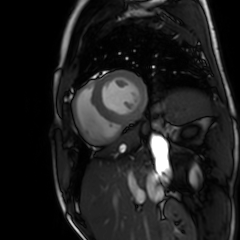};

\nextgroupplot[
scaled x ticks=manual:{}{\pgfmathparse{#1}},
scaled y ticks=manual:{}{\pgfmathparse{#1}},
width=\imgwidth,
height=\imgwidth,
tick align=outside,
x grid style={darkgray176},
xmajorticks=false,
xmin=0, xmax=200,
xtick style={color=black},
xticklabels={},
y dir=reverse,
y grid style={darkgray176},
ymajorticks=false,
ymin=0, ymax=200,
ytick style={color=black},
yticklabels={},
ylabel={$\frameindex = 116$}
]
\addplot graphics [includegraphics cmd=\pgfimage,xmin=0, xmax=200, ymin=200, ymax=0] {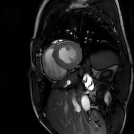};

\nextgroupplot[
scaled x ticks=manual:{}{\pgfmathparse{#1}},
scaled y ticks=manual:{}{\pgfmathparse{#1}},
width=\imgwidth,
height=\imgwidth,
tick align=outside,
x grid style={darkgray176},
xmajorticks=false,
xmin=0, xmax=200,
xtick style={color=black},
xticklabels={},
y dir=reverse,
y grid style={darkgray176},
ymajorticks=false,
ymin=0, ymax=200,
ytick style={color=black},
yticklabels={},
]
\addplot graphics [includegraphics cmd=\pgfimage, xmin=0, xmax=200, ymin=200, ymax=0] {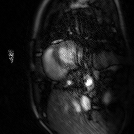};

\nextgroupplot[
scaled x ticks=manual:{}{\pgfmathparse{#1}},
scaled y ticks=manual:{}{\pgfmathparse{#1}},
width=\imgwidth,
height=\imgwidth,
tick align=outside,
x grid style={darkgray176},
xmajorticks=false,
xmin=0, xmax=200,
xtick style={color=black},
xticklabels={},
y dir=reverse,
y grid style={darkgray176},
ymajorticks=false,
ymin=0, ymax=200,
ytick style={color=black},
yticklabels={}
]
\addplot graphics [includegraphics cmd=\pgfimage,xmin=0, xmax=200, ymin=200, ymax=0] {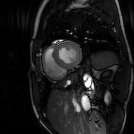};

\nextgroupplot[
scaled x ticks=manual:{}{\pgfmathparse{#1}},
scaled y ticks=manual:{}{\pgfmathparse{#1}},
width=\imgwidth,
height=\imgwidth,
tick align=outside,
x grid style={darkgray176},
xmajorticks=false,
xmin=0, xmax=200,
xtick style={color=black},
xticklabels={},
y dir=reverse,
y grid style={darkgray176},
ymajorticks=false,
ymin=0, ymax=200,
ytick style={color=black},
yticklabels={}
]
\addplot graphics [includegraphics cmd=\pgfimage,xmin=0, xmax=200, ymin=200, ymax=0] {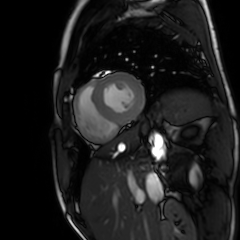};

\nextgroupplot[
scaled x ticks=manual:{}{\pgfmathparse{#1}},
scaled y ticks=manual:{}{\pgfmathparse{#1}},
width=\imgwidth,
height=\imgwidth,
tick align=outside,
x grid style={darkgray176},
xmajorticks=false,
xmin=0, xmax=200,
xtick style={color=black},
xticklabels={},
y dir=reverse,
y grid style={darkgray176},
ymajorticks=false,
ymin=0, ymax=200,
ytick style={color=black},
yticklabels={},
ylabel={$\frameindex = 126$}
]
\addplot graphics [includegraphics cmd=\pgfimage,xmin=0, xmax=200, ymin=200, ymax=0] {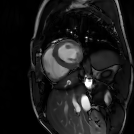};

\nextgroupplot[
scaled x ticks=manual:{}{\pgfmathparse{#1}},
scaled y ticks=manual:{}{\pgfmathparse{#1}},
width=\imgwidth,
height=\imgwidth,
tick align=outside,
x grid style={darkgray176},
xmajorticks=false,
xmin=0, xmax=200,
xtick style={color=black},
xticklabels={},
y dir=reverse,
y grid style={darkgray176},
ymajorticks=false,
ymin=0, ymax=200,
ytick style={color=black},
yticklabels={},
]
\addplot graphics [includegraphics cmd=\pgfimage, xmin=0, xmax=200, ymin=200, ymax=0] {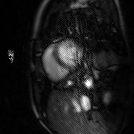};

\nextgroupplot[
scaled x ticks=manual:{}{\pgfmathparse{#1}},
scaled y ticks=manual:{}{\pgfmathparse{#1}},
width=\imgwidth,
height=\imgwidth,
tick align=outside,
x grid style={darkgray176},
xmajorticks=false,
xmin=0, xmax=200,
xtick style={color=black},
xticklabels={},
y dir=reverse,
y grid style={darkgray176},
ymajorticks=false,
ymin=0, ymax=200,
ytick style={color=black},
yticklabels={}
]
\addplot graphics [includegraphics cmd=\pgfimage,xmin=0, xmax=200, ymin=200, ymax=0] {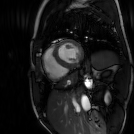};

\nextgroupplot[
scaled x ticks=manual:{}{\pgfmathparse{#1}},
scaled y ticks=manual:{}{\pgfmathparse{#1}},
width=\imgwidth,
height=\imgwidth,
tick align=outside,
x grid style={darkgray176},
xmajorticks=false,
xmin=0, xmax=200,
xtick style={color=black},
xticklabels={},
y dir=reverse,
y grid style={darkgray176},
ymajorticks=false,
ymin=0, ymax=200,
ytick style={color=black},
yticklabels={}
]
\addplot graphics [includegraphics cmd=\pgfimage,xmin=0, xmax=200, ymin=200, ymax=0] {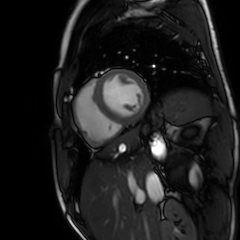};

\nextgroupplot[
scaled x ticks=manual:{}{\pgfmathparse{#1}},
scaled y ticks=manual:{}{\pgfmathparse{#1}},
width=\imgwidth,
height=\imgwidth,
tick align=outside,
x grid style={darkgray176},
xmajorticks=false,
xmin=0, xmax=200,
xtick style={color=black},
xticklabels={},
y dir=reverse,
y grid style={darkgray176},
ymajorticks=false,
ymin=0, ymax=200,
ytick style={color=black},
yticklabels={},
ylabel={$\frameindex = 136$}
]
\addplot graphics [includegraphics cmd=\pgfimage,xmin=0, xmax=200, ymin=200, ymax=0] {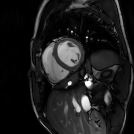};


\nextgroupplot[
scaled x ticks=manual:{}{\pgfmathparse{#1}},
scaled y ticks=manual:{}{\pgfmathparse{#1}},
width=\imgwidth,
height=\imgwidth,
tick align=outside,
x grid style={darkgray176},
xmajorticks=false,
xmin=0, xmax=200,
xtick style={color=black},
xticklabels={},
y dir=reverse,
y grid style={darkgray176},
ymajorticks=false,
ymin=0, ymax=200,
ytick style={color=black},
yticklabels={},
]
\addplot graphics [includegraphics cmd=\pgfimage, xmin=0, xmax=200, ymin=200, ymax=0] {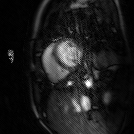};

\nextgroupplot[
scaled x ticks=manual:{}{\pgfmathparse{#1}},
scaled y ticks=manual:{}{\pgfmathparse{#1}},
width=\imgwidth,
height=\imgwidth,
tick align=outside,
x grid style={darkgray176},
xmajorticks=false,
xmin=0, xmax=200,
xtick style={color=black},
xticklabels={},
y dir=reverse,
y grid style={darkgray176},
ymajorticks=false,
ymin=0, ymax=200,
ytick style={color=black},
yticklabels={}
]
\addplot graphics [includegraphics cmd=\pgfimage,xmin=0, xmax=200, ymin=200, ymax=0] {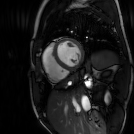};

\nextgroupplot[
scaled x ticks=manual:{}{\pgfmathparse{#1}},
scaled y ticks=manual:{}{\pgfmathparse{#1}},
width=\imgwidth,
height=\imgwidth,
tick align=outside,
x grid style={darkgray176},
xmajorticks=false,
xmin=0, xmax=200,
xtick style={color=black},
xticklabels={},
y dir=reverse,
y grid style={darkgray176},
ymajorticks=false,
ymin=0, ymax=200,
ytick style={color=black},
yticklabels={}
]
\addplot graphics [includegraphics cmd=\pgfimage,xmin=0, xmax=200, ymin=200, ymax=0] {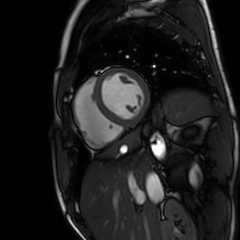};

\end{groupplot}

\end{tikzpicture}
    \vspace{1em} 
    \caption{The reconstructions of the low-resolution high-SNR dataset are shown with a larger field-of-view. It can be seen that the FMLP and the t-DIP achieve a similar reconstruction quality whereby the FMLP recovers small anatomic details in the heart, such as the papillary muscles, more accurately. The KFMLP, by contrast, suffers from aliasing-like artifacts.}
    \label{fig:reconstructions-10-full-res}
\end{figure}

\begin{figure}[H]
    \centering
\begin{tikzpicture}

\newcommand\imgwidth[0]{5cm}

\definecolor{darkgray176}{RGB}{176,176,176}

\begin{groupplot}[
  group style={
    group size=4 by 4,
    ylabels at=edge left,
    xlabels at=edge top,
    vertical sep=0.1cm,
    horizontal sep=0.1cm
  }]

\nextgroupplot[
scaled x ticks=manual:{}{\pgfmathparse{#1}},
scaled y ticks=manual:{}{\pgfmathparse{#1}},
width=\imgwidth,
height=\imgwidth,
tick align=outside,
x grid style={darkgray176},
xmajorticks=false,
xmin=0, xmax=200,
xtick style={color=black},
xticklabels={},
y dir=reverse,
y grid style={darkgray176},
ymajorticks=false,
ymin=0, ymax=200,
ytick style={color=black},
yticklabels={},
ylabel={$\frameindex = 106$},
xlabel={FMLP}
]
\addplot graphics [includegraphics cmd=\pgfimage, xmin=0, xmax=200, ymin=200, ymax=0] {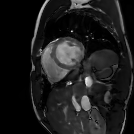};

\nextgroupplot[
scaled x ticks=manual:{}{\pgfmathparse{#1}},
scaled y ticks=manual:{}{\pgfmathparse{#1}},
width=\imgwidth,
height=\imgwidth,
tick align=outside,
x grid style={darkgray176},
xmajorticks=false,
xmin=0, xmax=200,
xtick style={color=black},
xticklabels={},
y dir=reverse,
y grid style={darkgray176},
ymajorticks=false,
ymin=0, ymax=200,
ytick style={color=black},
xlabel={KFMLP},
yticklabels={},
]
\addplot graphics [includegraphics cmd=\pgfimage, xmin=0, xmax=200, ymin=200, ymax=0] {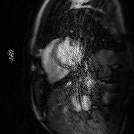};

\nextgroupplot[
scaled x ticks=manual:{}{\pgfmathparse{#1}},
scaled y ticks=manual:{}{\pgfmathparse{#1}},
width=\imgwidth,
height=\imgwidth,
tick align=outside,
x grid style={darkgray176},
xmajorticks=false,
xmin=0, xmax=200,
xtick style={color=black},
xticklabels={},
y dir=reverse,
y grid style={darkgray176},
ymajorticks=false,
ymin=0, ymax=200,
ytick style={color=black},
yticklabels={},
xlabel={t-DIP}
]
\addplot graphics [includegraphics cmd=\pgfimage,xmin=0, xmax=200, ymin=200, ymax=0] {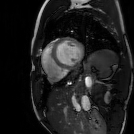};

\nextgroupplot[
scaled x ticks=manual:{}{\pgfmathparse{#1}},
scaled y ticks=manual:{}{\pgfmathparse{#1}},
width=\imgwidth,
height=\imgwidth,
tick align=outside,
x grid style={darkgray176},
xmajorticks=false,
xmin=0, xmax=200,
xtick style={color=black},
xticklabels={},
y dir=reverse,
y grid style={darkgray176},
ymajorticks=false,
ymin=0, ymax=200,
ytick style={color=black},
yticklabels={},
xlabel={BH reference}
]
\addplot graphics [includegraphics cmd=\pgfimage,xmin=0, xmax=200, ymin=200, ymax=0] {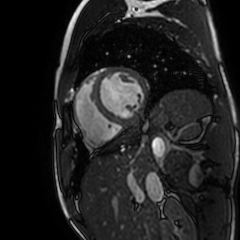};

\nextgroupplot[
scaled x ticks=manual:{}{\pgfmathparse{#1}},
scaled y ticks=manual:{}{\pgfmathparse{#1}},
width=\imgwidth,
height=\imgwidth,
tick align=outside,
x grid style={darkgray176},
xmajorticks=false,
xmin=0, xmax=200,
xtick style={color=black},
xticklabels={},
y dir=reverse,
y grid style={darkgray176},
ymajorticks=false,
ymin=0, ymax=200,
ytick style={color=black},
yticklabels={},
ylabel={$\frameindex = 116$}
]
\addplot graphics [includegraphics cmd=\pgfimage,xmin=0, xmax=200, ymin=200, ymax=0] {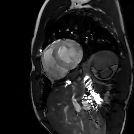};

\nextgroupplot[
scaled x ticks=manual:{}{\pgfmathparse{#1}},
scaled y ticks=manual:{}{\pgfmathparse{#1}},
width=\imgwidth,
height=\imgwidth,
tick align=outside,
x grid style={darkgray176},
xmajorticks=false,
xmin=0, xmax=200,
xtick style={color=black},
xticklabels={},
y dir=reverse,
y grid style={darkgray176},
ymajorticks=false,
ymin=0, ymax=200,
ytick style={color=black},
yticklabels={},
]
\addplot graphics [includegraphics cmd=\pgfimage, xmin=0, xmax=200, ymin=200, ymax=0] {figures/include/fmlp-tdip-bh-15-225/kfmlp-000.png};

\nextgroupplot[
scaled x ticks=manual:{}{\pgfmathparse{#1}},
scaled y ticks=manual:{}{\pgfmathparse{#1}},
width=\imgwidth,
height=\imgwidth,
tick align=outside,
x grid style={darkgray176},
xmajorticks=false,
xmin=0, xmax=200,
xtick style={color=black},
xticklabels={},
y dir=reverse,
y grid style={darkgray176},
ymajorticks=false,
ymin=0, ymax=200,
ytick style={color=black},
yticklabels={}
]
\addplot graphics [includegraphics cmd=\pgfimage,xmin=0, xmax=200, ymin=200, ymax=0] {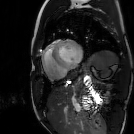};

\nextgroupplot[
scaled x ticks=manual:{}{\pgfmathparse{#1}},
scaled y ticks=manual:{}{\pgfmathparse{#1}},
width=\imgwidth,
height=\imgwidth,
tick align=outside,
x grid style={darkgray176},
xmajorticks=false,
xmin=0, xmax=200,
xtick style={color=black},
xticklabels={},
y dir=reverse,
y grid style={darkgray176},
ymajorticks=false,
ymin=0, ymax=200,
ytick style={color=black},
yticklabels={}
]
\addplot graphics [includegraphics cmd=\pgfimage,xmin=0, xmax=200, ymin=200, ymax=0] {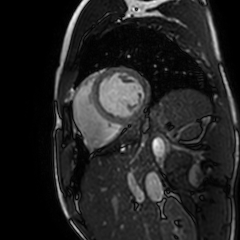};

\nextgroupplot[
scaled x ticks=manual:{}{\pgfmathparse{#1}},
scaled y ticks=manual:{}{\pgfmathparse{#1}},
width=\imgwidth,
height=\imgwidth,
tick align=outside,
x grid style={darkgray176},
xmajorticks=false,
xmin=0, xmax=200,
xtick style={color=black},
xticklabels={},
y dir=reverse,
y grid style={darkgray176},
ymajorticks=false,
ymin=0, ymax=200,
ytick style={color=black},
yticklabels={},
ylabel={$\frameindex = 126$}
]
\addplot graphics [includegraphics cmd=\pgfimage,xmin=0, xmax=200, ymin=200, ymax=0] {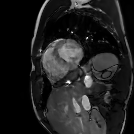};

\nextgroupplot[
scaled x ticks=manual:{}{\pgfmathparse{#1}},
scaled y ticks=manual:{}{\pgfmathparse{#1}},
width=\imgwidth,
height=\imgwidth,
tick align=outside,
x grid style={darkgray176},
xmajorticks=false,
xmin=0, xmax=200,
xtick style={color=black},
xticklabels={},
y dir=reverse,
y grid style={darkgray176},
ymajorticks=false,
ymin=0, ymax=200,
ytick style={color=black},
yticklabels={},
]
\addplot graphics [includegraphics cmd=\pgfimage, xmin=0, xmax=200, ymin=200, ymax=0] {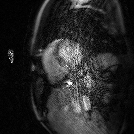};

\nextgroupplot[
scaled x ticks=manual:{}{\pgfmathparse{#1}},
scaled y ticks=manual:{}{\pgfmathparse{#1}},
width=\imgwidth,
height=\imgwidth,
tick align=outside,
x grid style={darkgray176},
xmajorticks=false,
xmin=0, xmax=200,
xtick style={color=black},
xticklabels={},
y dir=reverse,
y grid style={darkgray176},
ymajorticks=false,
ymin=0, ymax=200,
ytick style={color=black},
yticklabels={}
]
\addplot graphics [includegraphics cmd=\pgfimage,xmin=0, xmax=200, ymin=200, ymax=0] {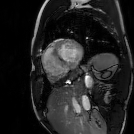};

\nextgroupplot[
scaled x ticks=manual:{}{\pgfmathparse{#1}},
scaled y ticks=manual:{}{\pgfmathparse{#1}},
width=\imgwidth,
height=\imgwidth,
tick align=outside,
x grid style={darkgray176},
xmajorticks=false,
xmin=0, xmax=200,
xtick style={color=black},
xticklabels={},
y dir=reverse,
y grid style={darkgray176},
ymajorticks=false,
ymin=0, ymax=200,
ytick style={color=black},
yticklabels={}
]
\addplot graphics [includegraphics cmd=\pgfimage,xmin=0, xmax=200, ymin=200, ymax=0] {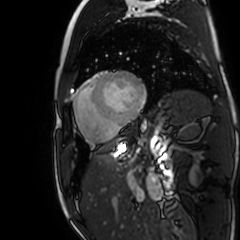};

\nextgroupplot[
scaled x ticks=manual:{}{\pgfmathparse{#1}},
scaled y ticks=manual:{}{\pgfmathparse{#1}},
width=\imgwidth,
height=\imgwidth,
tick align=outside,
x grid style={darkgray176},
xmajorticks=false,
xmin=0, xmax=200,
xtick style={color=black},
xticklabels={},
y dir=reverse,
y grid style={darkgray176},
ymajorticks=false,
ymin=0, ymax=200,
ytick style={color=black},
yticklabels={},
ylabel={$\frameindex = 136$}
]
\addplot graphics [includegraphics cmd=\pgfimage,xmin=0, xmax=200, ymin=200, ymax=0] {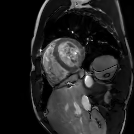};

\nextgroupplot[
scaled x ticks=manual:{}{\pgfmathparse{#1}},
scaled y ticks=manual:{}{\pgfmathparse{#1}},
width=\imgwidth,
height=\imgwidth,
tick align=outside,
x grid style={darkgray176},
xmajorticks=false,
xmin=0, xmax=200,
xtick style={color=black},
xticklabels={},
y dir=reverse,
y grid style={darkgray176},
ymajorticks=false,
ymin=0, ymax=200,
ytick style={color=black},
yticklabels={},
]
\addplot graphics [includegraphics cmd=\pgfimage, xmin=0, xmax=200, ymin=200, ymax=0] {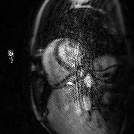};

\nextgroupplot[
scaled x ticks=manual:{}{\pgfmathparse{#1}},
scaled y ticks=manual:{}{\pgfmathparse{#1}},
width=\imgwidth,
height=\imgwidth,
tick align=outside,
x grid style={darkgray176},
xmajorticks=false,
xmin=0, xmax=200,
xtick style={color=black},
xticklabels={},
y dir=reverse,
y grid style={darkgray176},
ymajorticks=false,
ymin=0, ymax=200,
ytick style={color=black},
yticklabels={}
]
\addplot graphics [includegraphics cmd=\pgfimage,xmin=0, xmax=200, ymin=200, ymax=0] {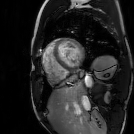};

\nextgroupplot[
scaled x ticks=manual:{}{\pgfmathparse{#1}},
scaled y ticks=manual:{}{\pgfmathparse{#1}},
width=\imgwidth,
height=\imgwidth,
tick align=outside,
x grid style={darkgray176},
xmajorticks=false,
xmin=0, xmax=200,
xtick style={color=black},
xticklabels={},
y dir=reverse,
y grid style={darkgray176},
ymajorticks=false,
ymin=0, ymax=200,
ytick style={color=black},
yticklabels={}
]
\addplot graphics [includegraphics cmd=\pgfimage,xmin=0, xmax=200, ymin=200, ymax=0] {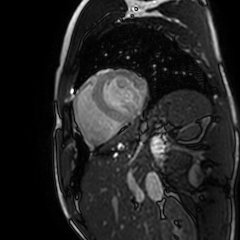};

\end{groupplot}

\end{tikzpicture}
    \vspace{1em} 
    \caption{The reconstructions of the low-resolution low-SNR dataset are shown with a larger field-of-view. The figure shows that the FMLP and the t-DIP exhibit similar artifacts and are on par in terms of image quality. The KFMLP, however, has a low image quality with aliasing-like artifacts and noise superimposed on the image. The artifacts are most prominent in the center of the image around the heart where the motion is strongest. }
    \label{fig:reconstructions-15-full-res}
\end{figure}

\begin{figure}[H]
    \centering
\begin{tikzpicture}

\newcommand\imgwidth[0]{5cm}

\definecolor{darkgray176}{RGB}{176,176,176}

\begin{groupplot}[
  group style={
    group size=4 by 4,
    ylabels at=edge left,
    xlabels at=edge top,
    vertical sep=0.1cm,
    horizontal sep=0.1cm
  }]

\nextgroupplot[
scaled x ticks=manual:{}{\pgfmathparse{#1}},
scaled y ticks=manual:{}{\pgfmathparse{#1}},
width=\imgwidth,
height=\imgwidth,
tick align=outside,
x grid style={darkgray176},
xmajorticks=false,
xmin=0, xmax=200,
xtick style={color=black},
xticklabels={},
y dir=reverse,
y grid style={darkgray176},
ymajorticks=false,
ymin=0, ymax=200,
ytick style={color=black},
yticklabels={},
ylabel={$\frameindex = 106$},
xlabel={FMLP}
]
\addplot graphics [includegraphics cmd=\pgfimage, xmin=0, xmax=200, ymin=200, ymax=0] {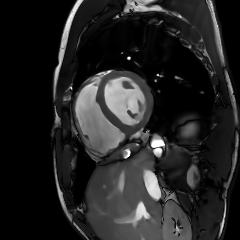};

\nextgroupplot[
scaled x ticks=manual:{}{\pgfmathparse{#1}},
scaled y ticks=manual:{}{\pgfmathparse{#1}},
width=\imgwidth,
height=\imgwidth,
tick align=outside,
x grid style={darkgray176},
xmajorticks=false,
xmin=0, xmax=200,
xtick style={color=black},
xticklabels={},
y dir=reverse,
y grid style={darkgray176},
ymajorticks=false,
ymin=0, ymax=200,
ytick style={color=black},
xlabel={KFMLP},
yticklabels={},
]
\addplot graphics [includegraphics cmd=\pgfimage, xmin=0, xmax=200, ymin=200, ymax=0] {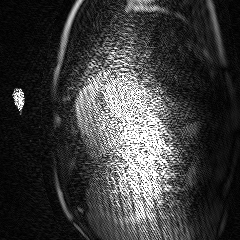};

\nextgroupplot[
scaled x ticks=manual:{}{\pgfmathparse{#1}},
scaled y ticks=manual:{}{\pgfmathparse{#1}},
width=\imgwidth,
height=\imgwidth,
tick align=outside,
x grid style={darkgray176},
xmajorticks=false,
xmin=0, xmax=200,
xtick style={color=black},
xticklabels={},
y dir=reverse,
y grid style={darkgray176},
ymajorticks=false,
ymin=0, ymax=200,
ytick style={color=black},
yticklabels={},
xlabel={t-DIP}
]
\addplot graphics [includegraphics cmd=\pgfimage,xmin=0, xmax=200, ymin=200, ymax=0] {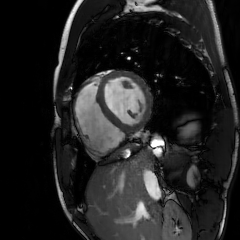};

\nextgroupplot[
scaled x ticks=manual:{}{\pgfmathparse{#1}},
scaled y ticks=manual:{}{\pgfmathparse{#1}},
width=\imgwidth,
height=\imgwidth,
tick align=outside,
x grid style={darkgray176},
xmajorticks=false,
xmin=0, xmax=200,
xtick style={color=black},
xticklabels={},
y dir=reverse,
y grid style={darkgray176},
ymajorticks=false,
ymin=0, ymax=200,
ytick style={color=black},
yticklabels={},
xlabel={BH reference}
]
\addplot graphics [includegraphics cmd=\pgfimage,xmin=0, xmax=200, ymin=200, ymax=0] {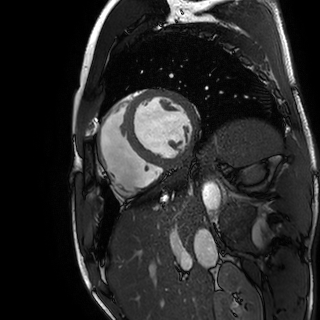};

\nextgroupplot[
scaled x ticks=manual:{}{\pgfmathparse{#1}},
scaled y ticks=manual:{}{\pgfmathparse{#1}},
width=\imgwidth,
height=\imgwidth,
tick align=outside,
x grid style={darkgray176},
xmajorticks=false,
xmin=0, xmax=200,
xtick style={color=black},
xticklabels={},
y dir=reverse,
y grid style={darkgray176},
ymajorticks=false,
ymin=0, ymax=200,
ytick style={color=black},
yticklabels={},
ylabel={$\frameindex = 116$}
]
\addplot graphics [includegraphics cmd=\pgfimage,xmin=0, xmax=200, ymin=200, ymax=0] {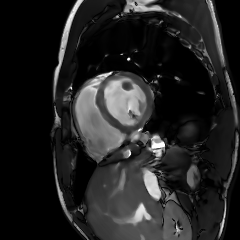};

\nextgroupplot[
scaled x ticks=manual:{}{\pgfmathparse{#1}},
scaled y ticks=manual:{}{\pgfmathparse{#1}},
width=\imgwidth,
height=\imgwidth,
tick align=outside,
x grid style={darkgray176},
xmajorticks=false,
xmin=0, xmax=200,
xtick style={color=black},
xticklabels={},
y dir=reverse,
y grid style={darkgray176},
ymajorticks=false,
ymin=0, ymax=200,
ytick style={color=black},
yticklabels={},
]
\addplot graphics [includegraphics cmd=\pgfimage, xmin=0, xmax=200, ymin=200, ymax=0] {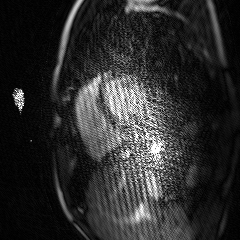};

\nextgroupplot[
scaled x ticks=manual:{}{\pgfmathparse{#1}},
scaled y ticks=manual:{}{\pgfmathparse{#1}},
width=\imgwidth,
height=\imgwidth,
tick align=outside,
x grid style={darkgray176},
xmajorticks=false,
xmin=0, xmax=200,
xtick style={color=black},
xticklabels={},
y dir=reverse,
y grid style={darkgray176},
ymajorticks=false,
ymin=0, ymax=200,
ytick style={color=black},
yticklabels={}
]
\addplot graphics [includegraphics cmd=\pgfimage,xmin=0, xmax=200, ymin=200, ymax=0] {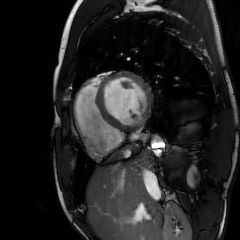};

\nextgroupplot[
scaled x ticks=manual:{}{\pgfmathparse{#1}},
scaled y ticks=manual:{}{\pgfmathparse{#1}},
width=\imgwidth,
height=\imgwidth,
tick align=outside,
x grid style={darkgray176},
xmajorticks=false,
xmin=0, xmax=200,
xtick style={color=black},
xticklabels={},
y dir=reverse,
y grid style={darkgray176},
ymajorticks=false,
ymin=0, ymax=200,
ytick style={color=black},
yticklabels={}
]
\addplot graphics [includegraphics cmd=\pgfimage,xmin=0, xmax=200, ymin=200, ymax=0] {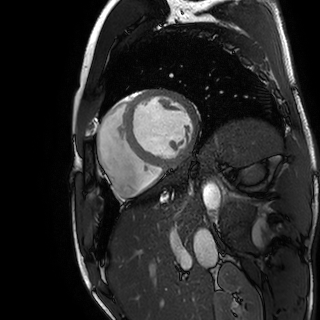};

\nextgroupplot[
scaled x ticks=manual:{}{\pgfmathparse{#1}},
scaled y ticks=manual:{}{\pgfmathparse{#1}},
width=\imgwidth,
height=\imgwidth,
tick align=outside,
x grid style={darkgray176},
xmajorticks=false,
xmin=0, xmax=200,
xtick style={color=black},
xticklabels={},
y dir=reverse,
y grid style={darkgray176},
ymajorticks=false,
ymin=0, ymax=200,
ytick style={color=black},
yticklabels={},
ylabel={$\frameindex = 126$}
]
\addplot graphics [includegraphics cmd=\pgfimage,xmin=0, xmax=200, ymin=200, ymax=0] {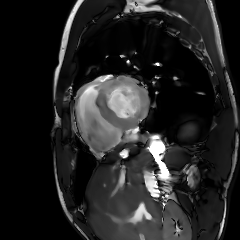};

\nextgroupplot[
scaled x ticks=manual:{}{\pgfmathparse{#1}},
scaled y ticks=manual:{}{\pgfmathparse{#1}},
width=\imgwidth,
height=\imgwidth,
tick align=outside,
x grid style={darkgray176},
xmajorticks=false,
xmin=0, xmax=200,
xtick style={color=black},
xticklabels={},
y dir=reverse,
y grid style={darkgray176},
ymajorticks=false,
ymin=0, ymax=200,
ytick style={color=black},
yticklabels={},
]
\addplot graphics [includegraphics cmd=\pgfimage, xmin=0, xmax=200, ymin=200, ymax=0] {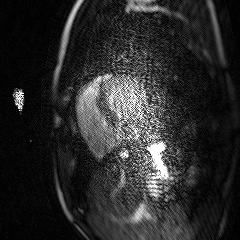};

\nextgroupplot[
scaled x ticks=manual:{}{\pgfmathparse{#1}},
scaled y ticks=manual:{}{\pgfmathparse{#1}},
width=\imgwidth,
height=\imgwidth,
tick align=outside,
x grid style={darkgray176},
xmajorticks=false,
xmin=0, xmax=200,
xtick style={color=black},
xticklabels={},
y dir=reverse,
y grid style={darkgray176},
ymajorticks=false,
ymin=0, ymax=200,
ytick style={color=black},
yticklabels={}
]
\addplot graphics [includegraphics cmd=\pgfimage,xmin=0, xmax=200, ymin=200, ymax=0] {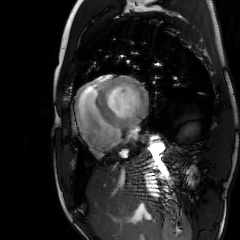};

\nextgroupplot[
scaled x ticks=manual:{}{\pgfmathparse{#1}},
scaled y ticks=manual:{}{\pgfmathparse{#1}},
width=\imgwidth,
height=\imgwidth,
tick align=outside,
x grid style={darkgray176},
xmajorticks=false,
xmin=0, xmax=200,
xtick style={color=black},
xticklabels={},
y dir=reverse,
y grid style={darkgray176},
ymajorticks=false,
ymin=0, ymax=200,
ytick style={color=black},
yticklabels={}
]
\addplot graphics [includegraphics cmd=\pgfimage,xmin=0, xmax=200, ymin=200, ymax=0] {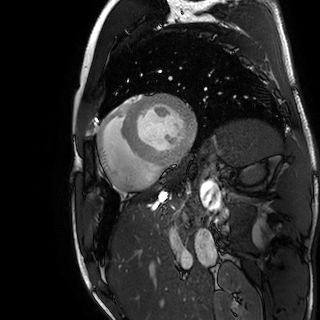};

\nextgroupplot[
scaled x ticks=manual:{}{\pgfmathparse{#1}},
scaled y ticks=manual:{}{\pgfmathparse{#1}},
width=\imgwidth,
height=\imgwidth,
tick align=outside,
x grid style={darkgray176},
xmajorticks=false,
xmin=0, xmax=200,
xtick style={color=black},
xticklabels={},
y dir=reverse,
y grid style={darkgray176},
ymajorticks=false,
ymin=0, ymax=200,
ytick style={color=black},
yticklabels={},
ylabel={$\frameindex = 136$}
]
\addplot graphics [includegraphics cmd=\pgfimage,xmin=0, xmax=200, ymin=200, ymax=0] {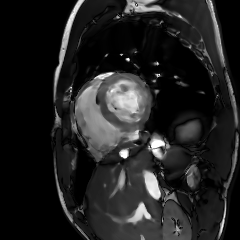};

\nextgroupplot[
scaled x ticks=manual:{}{\pgfmathparse{#1}},
scaled y ticks=manual:{}{\pgfmathparse{#1}},
width=\imgwidth,
height=\imgwidth,
tick align=outside,
x grid style={darkgray176},
xmajorticks=false,
xmin=0, xmax=200,
xtick style={color=black},
xticklabels={},
y dir=reverse,
y grid style={darkgray176},
ymajorticks=false,
ymin=0, ymax=200,
ytick style={color=black},
yticklabels={},
]
\addplot graphics [includegraphics cmd=\pgfimage, xmin=0, xmax=200, ymin=200, ymax=0] {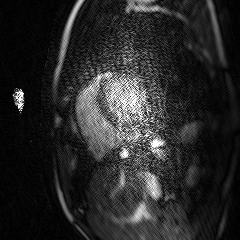};

\nextgroupplot[
scaled x ticks=manual:{}{\pgfmathparse{#1}},
scaled y ticks=manual:{}{\pgfmathparse{#1}},
width=\imgwidth,
height=\imgwidth,
tick align=outside,
x grid style={darkgray176},
xmajorticks=false,
xmin=0, xmax=200,
xtick style={color=black},
xticklabels={},
y dir=reverse,
y grid style={darkgray176},
ymajorticks=false,
ymin=0, ymax=200,
ytick style={color=black},
yticklabels={}
]
\addplot graphics [includegraphics cmd=\pgfimage,xmin=0, xmax=200, ymin=200, ymax=0] {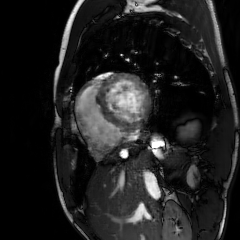};

\nextgroupplot[
scaled x ticks=manual:{}{\pgfmathparse{#1}},
scaled y ticks=manual:{}{\pgfmathparse{#1}},
width=\imgwidth,
height=\imgwidth,
tick align=outside,
x grid style={darkgray176},
xmajorticks=false,
xmin=0, xmax=200,
xtick style={color=black},
xticklabels={},
y dir=reverse,
y grid style={darkgray176},
ymajorticks=false,
ymin=0, ymax=200,
ytick style={color=black},
yticklabels={}
]
\addplot graphics [includegraphics cmd=\pgfimage,xmin=0, xmax=200, ymin=200, ymax=0] {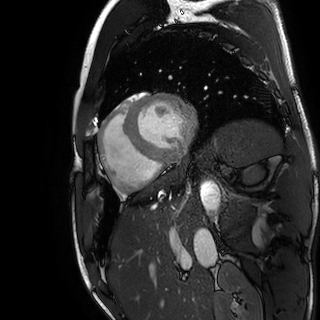};

\end{groupplot}

\end{tikzpicture}
    \vspace{1em} 
    \caption{The reconstructions of the high-resolution dataset are shown with a larger field-of-view. The image quality of the FMLP and the t-DIP are similar, whereas the KFMLP's reconstructions are distorted by noise that obscures meaningful anatomic details in the heart.}
    \label{fig:reconstructions-20-full-res}
\end{figure}

The reconstructions shown in the main document are truncated to the region-of-interest, the heart. For completeness, the reconstructions are shown with a larger field-of-view in Figures~\ref{fig:reconstructions-10-full-res}, \ref{fig:reconstructions-15-full-res}, and \ref{fig:reconstructions-20-full-res} for the low-resolution high-SNR, the low-resolution low-SNR, and the high-resolution dataset, respectively.

\FloatBarrier
\section{Measurement Parameters of the Datasets}
\label{sec:dataset-details}
\begin{figure}[h]
  \centering
  \input{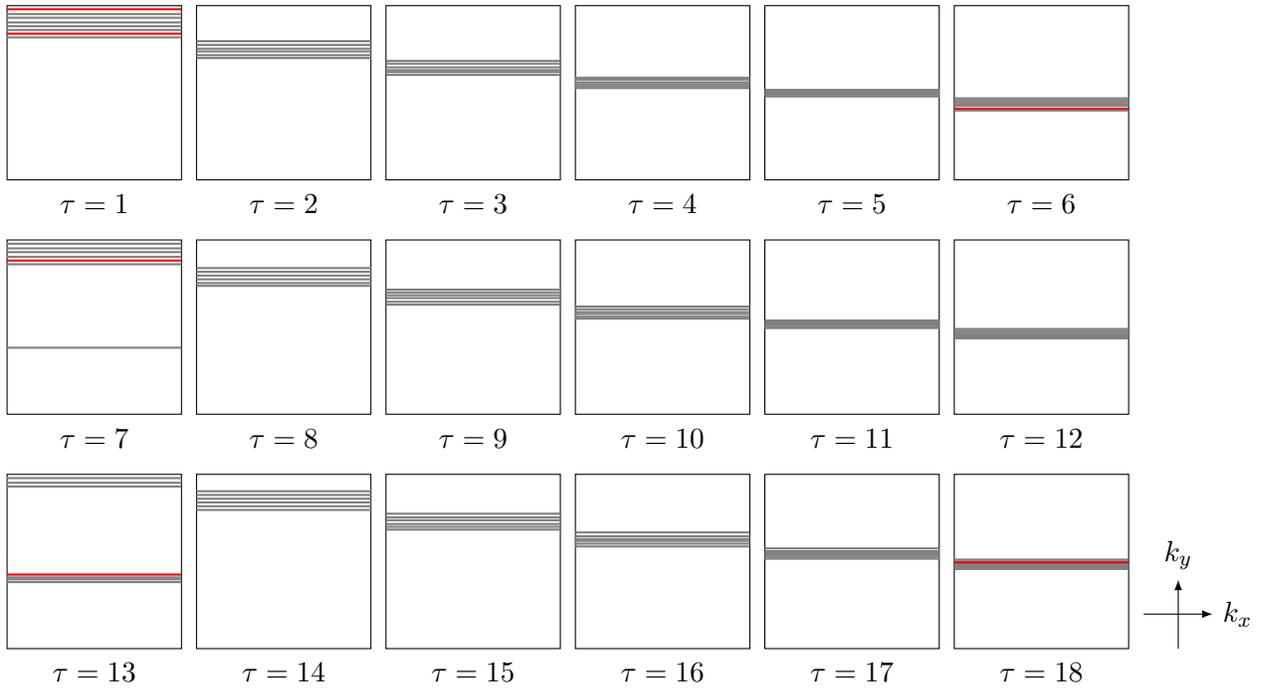}
  \caption{The sampling masks of the first 18 frames are illustrated for the low-resolution datasets with a binning into six lines per frame. The six $\ky$-lines that are used for training are colored in gray and the randomly selected validation lines are colored in red. The $\ky$-lines lines are measured in descending order from positive to negative $\ky$-coordinates. Due to the sampling pattern, validation lines are always in proximity of lines that are part of the training dataset.}
  \label{fig:sampling-pattern}
\end{figure}

The experimental datasets were acquired with a 2D balanced steady-state free precession (SSFP) sequence using a flip angle of \SI{45}{\degree} on a \SI{3}{\tesla} Elition X scanner (Philips Healthcare, The Netherlands). The free-breathing (FB) scans are taken with $C=25$ active receiver coils, while the breath-hold (BH) scans use a total of $C=26$ receiver coils.
The free-breathing datasets are recorded with a partial-Fourier deterministic sampling pattern that covers \SI{62.5}{\percent} of the k-space in phase-encoding ($\ky$) direction. The pattern of the low-resolution datasets is illustrated in Figure~\ref{fig:sampling-pattern}.
Further parameters of the free-breathing datasets are listed in Table~\ref{tab:fb-scan-parameters}.

The breath-hold scans are sampled using a SENSE sampling pattern with an acceleration factor of $R=2$ and 40 dynamics per cardiac cycle. For further details see Table~\ref{tab:bh-scan-parameters}.

\begin{table}[h]
    \centering
    \begin{tabular}{c|c|c|c}
    \specialrule{.15em}{.05em}{.05em}
    \shortstack{\strut Experiment     \\ ~ }                               &  \shortstack{\strut low resolution, \\  high-SNR}      &  \shortstack{\strut low resolution, \\  low-SNR}  & \shortstack{\strut high resolution \\ ~}         \\
        \midrule
        Slice thickness [\SI{}{\milli\metre}]                &  10 &  2.25 &  5 \\
        Field-of-view [\SI{}{\milli\metre^2}]                & 600 x 420 & 600 x 420 & 600 x 420 \\
        Acquisition matrix size                              & 264 x 186 & 264 x 186 & 480 x 334\\
        TR / TE [\SI{}{\milli\second}]                       & 2.8 / 1.39 & 3.5 / 1.76 & 3.4 / 1.71\\
        \specialrule{.15em}{.05em}{.05em}
    \end{tabular}
    \vspace{0.5em}
    \caption{Experiment-specific sequence parameters of the free-breathing scans.}
    \label{tab:fb-scan-parameters}
\end{table}

\begin{table}[h]
    \centering
    \setlength{\tabcolsep}{4pt}
    \begin{tabular}{c|c|c|c}
    \specialrule{.15em}{.05em}{.05em}
    \shortstack{\strut Experiment     \\ ~ }                               &  \shortstack{\strut low resolution, \\  high-SNR}      &  \shortstack{\strut low resolution, \\  low-SNR}  & \shortstack{\strut high resolution \\ ~}         \\
        \midrule
        Slice thickness [\SI{}{\milli\metre}]                &  10          &  2.25          &  5           \\
        Field-of-view [\SI{}{\milli\metre^2}]                & 600 x 540   & 600 x 540   & 600 x 330\\
        Acquisition matrix size                              & 264 x 240   & 264 x 240   & 480 x 264\\
        TR / TE [\SI{}{\milli\second}]                       & 2.8 / 1.38  & 3.5 / 1.75   & 3.5 / 1.77 \\
        TFE factor / shots / interval [\SI{}{\milli\second}] & 16 / 5 / 44.2 & 12 / 6 / 41.9 & 12 / 11 / 42.4\\
        \specialrule{.15em}{.05em}{.05em}
    \end{tabular}
    \vspace{0.5em}
    \caption{Experiment-specific sequence parameters of the breath-hold scans.}
    \label{tab:bh-scan-parameters}
\end{table}

\end{appendices}

\end{document}